\documentclass[manuscript]{aastex}
\newcommand{\Lpar}{L_{*}}
\begin{document}
\title{The fate of sub-micron circumplanetary dust grains II: Multipolar Fields}
\author{Daniel Jontof-Hutter\altaffilmark{1} Douglas P. Hamilton\altaffilmark{1}}
\affil{Astronomy Department, University of Maryland, College Park, MD 20742-2421}

\begin{abstract}
We study the radial and vertical stability of dust grains launched
with all charge-to-mass ratios at arbitrary distances from rotating
planets with complex magnetic fields. We show that the aligned dipole
magnetic field model analyzed by \citet{jh12} is an excellent
approximation in most cases, but that fundamentally new physics arises
with the inclusion of non-axisymmetric magnetic field terms. In
particular, large numbers of distant negatively-charged dust grains,
stable in a magnetic dipole, can be driven to escape by a more complex
field. We trace the origin of the instability to overlapping Lorentz
resonances which are extremely powerful when the gravitational and
electromagnetic forces on a dust grain are comparable. These
resonances enable a dust grain to tap the spin energy of the planet to
power its escape. We also explore the relatively minor influence of
different launch speeds and the far more important effects of variable
grain charge. Only the latter are capable of significantly affecting
the micron-sized grains that dominate visible and infrared images of
faint dust rings. Finally, we present full stability maps for Earth,
Jupiter, Saturn, Uranus, and Neptune with magnetic fields modeled out
to octupole order. Not surprisingly, dust in the tortured magnetic
fields of Uranus and Neptune show the greatest instability.
\end{abstract}
\section{Introduction}
When Voyager 1 encountered Jupiter in 1979, the discovery of the
tenuous dusty ring system came as a complete surprise. Although the earlier
Pioneer missions found some hints of a ring system, many thought that dust close to Jupiter would rapidly spiral
in by gas drag \citep{owe79}. Voyager 2 confirmed the existence of the
ring, and found the tiny satellites Metis and Adrastea that orbit
inside the classical Roche limit and are the most likely source of
ring material.

High speed impacts with small moons, as well as unseen large parent
bodies, replenish the dusty rings with debris of all sizes. Similar
sources of material for Saturn's tenuous inner D ring have not been
found, though massive particles in narrow ringlets with enhanced local
densities could serve as these sources
(\citealt{sho96,hed07}). In both environments, dust ejected by impacts
from parent bodies have essentially collisionless trajectories. As
debris particles acquire electric charges through interactions with
the plasma environment and solar radiation, the smallest reach
significant charge-to-mass ratios and, as a consequence, experience
strong electromagnetic (EM) forces as they orbit through the magnetic
field of their host planet.
 
For grains smaller than $\sim 1 \mu$m in radius, the EM force exceeds
perturbations from large satellites, the planetary oblateness and
solar radiation pressure \citep{hbh92}. Even smaller dust grains may
have orbits that are immediately unstable to either radial motion if
the grains are positively-charged (\citealt{hb93a,hmg93b}), or
vertical motion (both positively- and negatively-charged:
\citealt{nh82}). Many authors have studied various aspects of
charged-particle dynamics. For a recent review, see \citet{jh12}, who
derived analytic stability boundaries for the idealized case of grains with
constant charge, launched at the Kepler speed in an aligned dipolar
planetary magnetic field. As in that study, the boundaries between
stable and unstable orbits are of particular interest to us
here; these depend on the launch distance from the planet and the
charge-to-mass ratio of an individual dust grain.

The aim of this paper is to explore the sensitivity of these stability
boundaries to more realistic situations. We relax the idealized
assumptions of \citet{jh12} above by considering i) non-zero ejecta speeds from the
parent body, ii) higher-order magnetic field components, and iii)
variable electric potentials on dust grains. We use Jupiter as our
model planet since it has both a complex multipolar magnetic field and
a well-studied dusty ring system
(\citealt{bur99,dep99,ock99,bro04,thr04,sho08,kru09}). After a detailed
study of Jupiter, we then present stability maps for motion in the
complex magnetic fields of Earth, Saturn, Uranus and Neptune. We begin
by recapping stability results for a simple dipolar planetary magnetic
field from \citet{jh12}.

\section{Motion in an Aligned Dipolar Magnetic Field}
The charge-to-mass ratio for Kepler-launched grains can be
conveniently described by the ratio of the force induced by the
corotational electric field of the planet with gravity, given by
\begin{equation}
\label{Lpar}
\Lpar = \frac{qg_{10} R_p^3 \Omega_p}{GM_pmc}
\end{equation}
(\citealt{ham93a,ham93b,jh12}). Here, $q$ and $m$ are the electric
charge and mass of a dust grain, $g_{10}$ is the dipolar magnetic
field strength at the equator, $R_p$ and $M_p$ are the planetary
radius and mass respectively, $\Omega_p$ is the spin frequency of the
planet, and $G$ and $c$ are the gravitational constant and speed of
light. As a dimensionless independent variable, $\Lpar$ accounts for
all the relevant planetary parameters and avoids undue focus on the
grain's size, shape, density and electric potential which are all
poorly constrained. The sign of $\Lpar$ depends on the product $q
g_{10}$, and its value can easily be converted to a grain radius $a_d$
for specified grain properties.

For large grains dominated by gravity and orbiting with semi-major axis
$a$ (the Kepler limit), azimuthal, radial, and vertical motions have the same frequency
\begin{equation}
n_c = \left(\frac{G M_p}{a^3}\right)^{\frac12},
\label{nkep}
\end{equation}
but for higher charge-to-mass ratios, these frequencies differ. As the
charge-to-mass ratio is raised, these frequencies slowly diverge from
one another \citep{ham93a}, and for an aligned dipolar magnetic field,
explicit expressions valid for all charge-to-mass ratios are available
\citep{jh12}. As these expressions will prove useful for our current
study, we reproduce them here.

General motions in this problem can be conveniently separated into
epicyclic motion about a guiding center which in turn circles the planet at an
azimuthal angular speed $\omega_c$. Where radial epicycles are small
on the scale of the grain's orbit, balancing the centrifugal force, the EM
force and gravity yields an expression for $\omega_c$:
\begin{equation}
0 =\omega_c^2 + \frac{G M_p \Lpar}{r_c^3}\left(1-\frac{\omega_c}{\Omega_p} \right) -\frac{G M_p}{r_c^3}.
\label{omegac}
\end{equation}
(\citealt{nh82,mhh03,jh12}).  Here and throughout, the subscript $c$
refers to the guiding center of motion. The distance to the guiding
center of motion, $r_c$, is just the semi-major axis $a$ in the Kepler
limit. Note that for gravity-dominated grains ($\Lpar \rightarrow 0$),
we have $\omega_c^2 = G M_p/r_c^3 = n_c^2$ in agreement with
Eq.~\ref{nkep}. In the strong EM limit ($|\Lpar| \rightarrow \pm \infty$), $\omega_c \rightarrow \Omega_p$ and the
grains are nearly locked to the magnetic field lines.

The radial or epicyclic frequency $\kappa_c$ satisfies
\begin{equation}
\kappa_c^2 = \omega_c^2-4\omega_c\Omega_{gc}+\Omega_{gc}^2,
\label{kappa}
\end{equation}
(\citealt{mhh82,mhh03,jh12}) at the guiding center, where $\Omega_{gc}
= qB/mc = n_c^2 \Lpar/\Omega_p$ is the frequency of gyromotion. In the EM-dominated
Lorentz regime, $\kappa_c = \Omega_{gc}$. In the gravity-dominated
Kepler regime, $\Omega_{gc}\rightarrow 0$ and $\kappa_c \rightarrow
n_c$, the Kepler orbital frequency, as expected.

Most grains are radially confined, suffering excursions of
\begin{equation}
r_g = \frac{r_L(\Omega_p-n_L)\Omega_{gL}}{\Omega_{gL}^2-\Omega_{gL}(3\Omega_p+n_L)+n_L^2}
\label{rg}
\end{equation}
where $r_g$, the gyroradius, is much smaller than $r_L$, the launch
distance. Here $n_L = \sqrt{G M_p/r_L^3}$ and $\Omega_{gL} =
n_L^2\Lpar/\Omega_p$ are the Kepler frequency and the
gyrofrequency as determined at the launch distance
(\citealt{sb94,jh12}). The epicyclic model fails only for
positively-charged grains with $\Lpar \sim 1$, where the denominator
in Eq.~\ref{rg} becomes very small (see \citealt{jh12}, Fig. 6b).

Finally, the vertical motion of grains with stable epicycles in the
equatorial plane has frequency $\Omega_b$, where
\begin{equation}
\Omega_b^2 = 3 \omega_c^2-2n_c^2 + \frac{r_g^2}{\rho_c^2}\left(\frac{9}{2}\Omega_{gc}^2-\frac{9}{2}\Omega_{gc}\dot{\phi}_c -\frac{3}{2}n_c^2 \right)
\label{Omb}
\end{equation}    
\citep{jh12}. Here, $\dot{\phi}_c = \omega_c-\Omega_p$ is the
azimuthal motion of the guiding center in the frame rotating with the
planet. Equation~\ref{Omb} is valid as long as $r_g/r_L << 1$ which holds in both the Kepler ($\Lpar \rightarrow 0$) and Lorentz ($\Lpar \rightarrow \pm \infty$) limits. In the Kepler limit, all three $r_g^2$ terms are negligible and $\Omega_b
\rightarrow n_c$, while in the Lorentz limit, only the last two terms can be ignored and $\Omega_b^2 \rightarrow \frac{15}{2}\Omega_p^2 -9\Omega_p
n_c+\frac{5}{2}n_c^2$.

Where $\Omega_b$ tends to zero, grains in the equatorial plane become locally vertically unstable. Equation~\ref{Omb} provides good
agreement with numerical data on the location and charge-to-mass ratio
of boundaries between vertically stable and unstable grains, with two
important caveats.

Firstly, in applying the epicyclic approximation, Eq.~\ref{Omb}
assumes that radial motions are very small on the scale of the orbit
($r_g << r_L$). In addition, Eq.~\ref{Omb} is averaged over one
gyrocycle, so the epicyclic motion must occur on a much shorter timescale
than any stable vertical oscillations ($\kappa_c >> \Omega_b$). Both
of these assumptions are easily met in the Lorentz limit, but both lose
accuracy as $\Lpar$ decreases, particularly for the positively-charged
grains which become radially unstable as $\Lpar \rightarrow 1$.

\begin{figure}[placement h]
\includegraphics [height = 2.1 in]{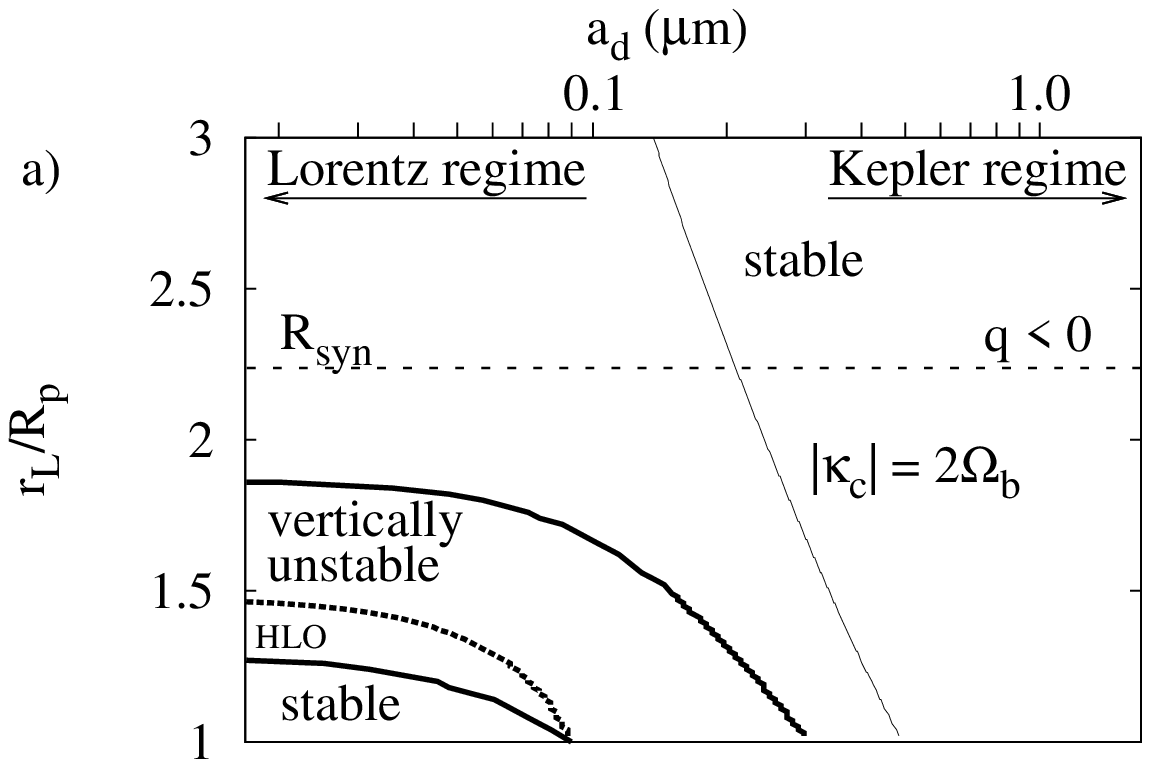}
\newline
\includegraphics [height = 2.1 in]{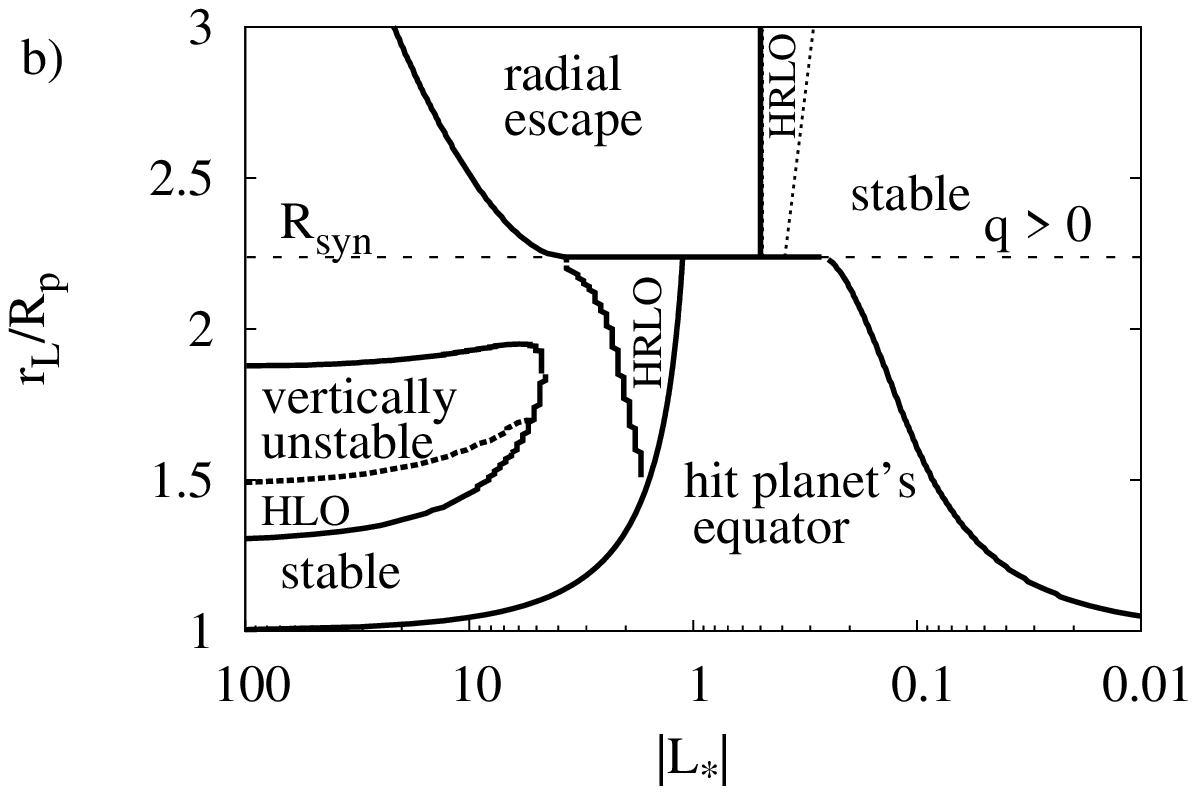}
\caption{Stability boundaries for a) negative ($q < 0$) and b)
  positive ($q > 0$) grains in an aligned dipole magnetic field
  for Jupiter. This figure summarizes numerical data from
  \citet{jh12}, with grain radii marked along the top axis
  corresponding to a $-5 V$ or $+5 V$ potential on a spherical grain
  of material density 1 g cm$^{-3}$. The vertically-unstable grains
  depart from the equatorial plane and climb to high latitudes
  immediately after launch, ultimately colliding with the
  planet. Directly below this unstable region are grain trajectories
  with high latitude oscillations (HLOs). The radially unstable grains
  ($q > 0$) escape if launched outside synchronous orbit
  ($R_{syn}$), or hit the planet if launched from within
  $R_{syn}$. Two regions of high radial and latitudinal oscillations (HRLOs) abut the radial instability region. Within $R_{syn}$,
  large inward radial excursions lead to vertical oscillations
  (roughly along magnetic field lines) which increase in amplitude
  until the grains strike the planet at high latitude. Outside
  $R_{syn}$, near $\Lpar =\frac12$, some grains experience HRLOs
  indefinitely. Finally, for $q < 0$, a curve traces grains that experience
  HRLOs following the 2:1 resonance between the epicyclic ($\kappa_c$)
  and vertical ($\Omega_b$) frequencies. }
\label{fig:1} 
\end{figure}

Secondly, setting $\Omega_b = 0$ determines local, as opposed to global, vertical stability in the equatorial plane of the spinning planet and its aligned dipolar
magnetic field. Local instability is a necessary condition for global
instability (whereby grains collide with the planet at high latitude),
but it is not always sufficient. High-latitude restoring forces often lead to stable high latitude oscillations (HLOs) \citep{jh12}. This class of orbits is more important for slow rotators
like the Earth than it is at Jupiter or
Saturn, but they do occur for the smallest grains inside $1.5 R_p$ at
Jupiter (Fig.~\ref{fig:1}).

With these two caveats in mind, we include the local and global
stability boundaries found from numerical integrations by \citet{jh12}
for Jupiter with an aligned dipolar magnetic field
model. Fig.~\ref{fig:1} highlights these regions for a range of
charge-to-mass ratios spanning four orders of magnitude and a suite of
launch distances from Jupiter's surface to beyond its synchronous
orbital distance, $R_{syn}$, with grains all launched at the local circular speed
of the large parent bodies. \citet{jh12} also derived analytic
approximations to most of the boundaries in Fig.~\ref{fig:1} (see
their Fig. 9).
\section{Jupiter}
We focus most of our attention on Jupiter as its magnetic field has
been well studied, and is known out to octupole order (we adopt the O4
model of \citealt{an76,des83}). The planet's magnetic field is
dominated by the dipolar terms: $g_{10}$ = 4.218 Gauss, $g_{11} =
-0.664$ Gauss, and $h_{11} = 0.264$ Gauss; these can be combined to
determine the dipole tilt angle: $\arctan
\left(\sqrt{(g_{11}^2+h_{11}^2)}/g_{10} \right) =
9.6^{\circ}$. The $g_{20}$ = -0.203 Gauss component
can be interpreted as a southward vertical offset to the dipole
field. Four additional quadrupolar and seven octupolar terms are
known, and the upcoming Juno mission will measure still higher-order
magnetic field coefficients for the first time. In this section, we
add various effects to a simple aligned dipole model to elucidate
their importance. We begin with non-zero launch speeds in the frame of
the parent particle, as likely occurs with impact ejecta.
\subsection{Varied Launch Speed}
\begin{figure}[placement h]
\includegraphics [height = 2.1 in]{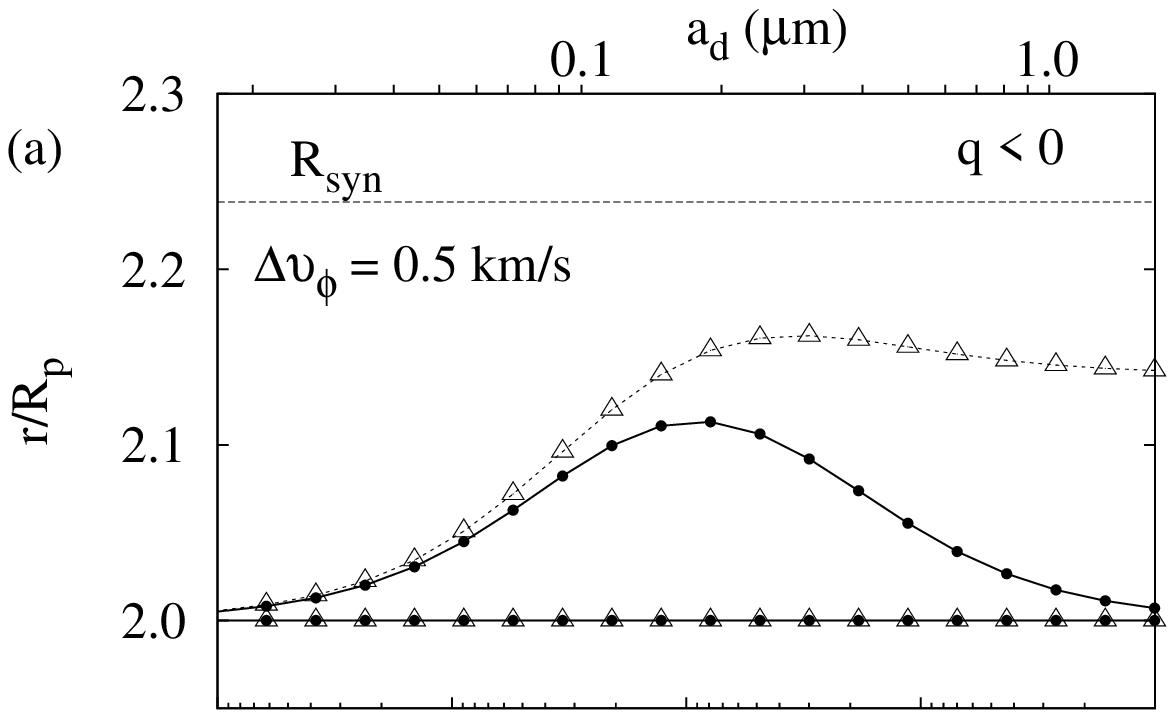}
\newline
\includegraphics [height = 2.1 in]{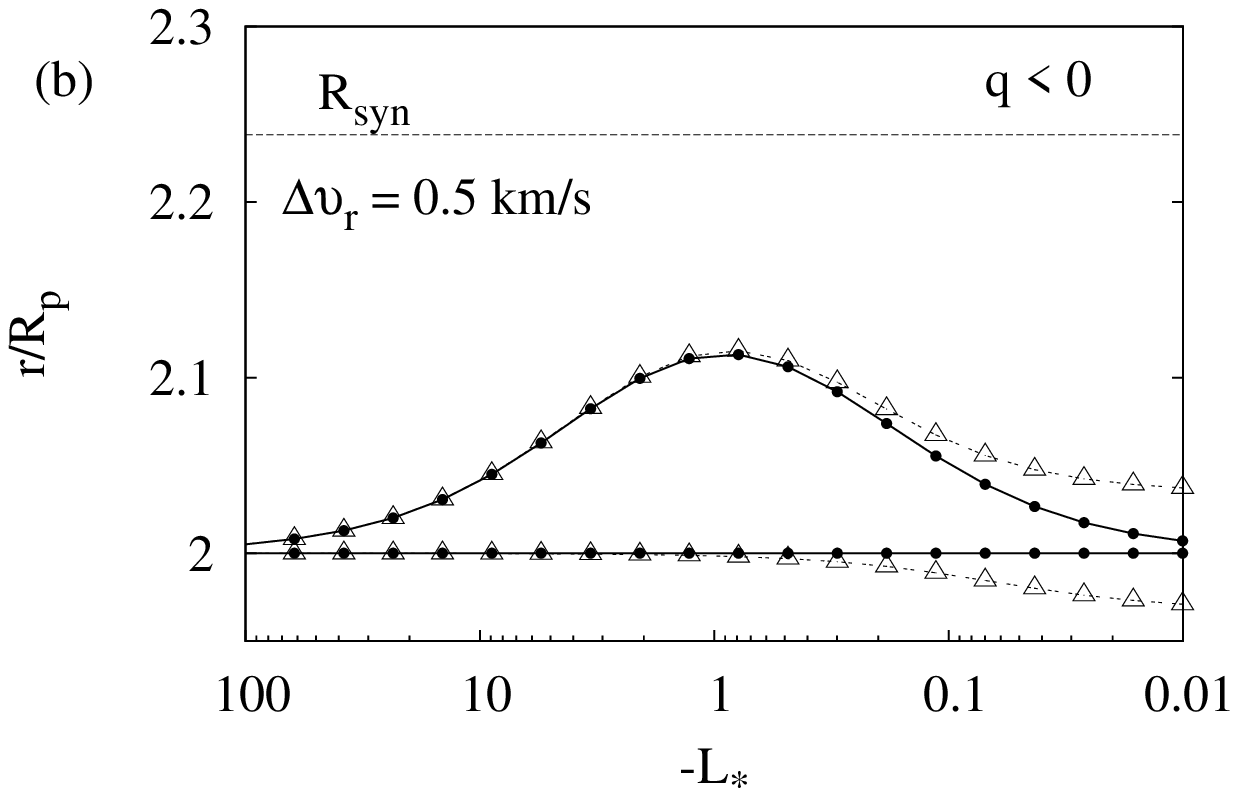}
\caption{Radial range of motion for negatively-charged grains on
  initially circular orbits subject to a) an azimuthal ($\Delta
  v_{\phi}$) and b) a radial ($\Delta v_r$) velocity impulse. The
  small filled circles indicate particles launched on circular orbits
  with $\Delta v$ = 0, while the large open triangles denote those
  launched with $\Delta v$ = 0.5 km s$^{-1}$. Synchronous orbit ($R_{syn}$) is indicated by a dashed line; the other lines simply track the data points.}
\label{fig:2} 
\end{figure}

Typical ejecta velocities from an impact are tens to hundreds of
meters per second in the rest frame of the parent body, in a cone
centered on the impact velocity vector \citep{dpl06}. Do these
non-circular launch speeds significantly affect the stability of charged dust
grains? To highlight the effect, we consider large initial velocities
of 0.5 km~s$^{-1}$ in the prograde azimuthal (Fig.~\ref{fig:2}a) and
radial (Fig.~\ref{fig:2}b) directions.
Even with these large speeds, we note that EM-dominated grains on the
left side of the plots are hardly affected. The Kepler speed at $r_L =
2.0 R_p$ is $v_k =$29.8 km s$^{-1}$, while the local magnetic field
lines rotate at $\Omega_p r_L$= 25.1 km s$^{-1}$. The azimuthal
impulse that we add, therefore, is only $\sim$10$\%$ of the
$\Omega_{gc}r_g$ = 4.7 km s$^{-1}$ gyrospeed and decreases the
gyroradius $r_g$ by a corresponding 10$\%$. Although important, this
effect is not noticeable on Fig.~\ref{fig:2}. 

On the other hand, grains
in the Kepler regime experience large radial excursions following a
launch impulse. For an azimuthal boost ($\Delta v_{\phi} > 0$), we can solve for the
semi-major axis $a$ and eccentricity $e$ from $r_L = a(1-e)$, and
\begin{equation}
v_k^2 = GM_p\left(\frac{2}{r}-\frac{1}{a}\right).
\label{visviva}
\end{equation} 
For $\Delta v_{\phi} << v_k$, the radial motions extend
outward from the launch position by $2ae \approx 4 r_L \Delta
  v_{\phi}/v_k \approx 0.134 R_p$ for the parameters of
Fig.~\ref{fig:2}a. Although the intermediate-sized grains have the
largest radial excursions in Fig.~\ref{fig:2}a, the grains in the
Kepler limit are most strongly affected by a $\Delta v_{\phi}$ kick.

Figure~\ref{fig:2}b shows that a radial impulse produces a more modest
radial range of motion than an azimuthal boost. In this case, the
impulse is perpendicular to the velocity of the parent body. For the
smallest grains, in the Lorentz limit, this has almost no effect on
the motion perpendicular to the field lines, and is akin to altering
the initial phase but not the size of the gyrocycle. As with the
azimuthal kick, a radial impulse has the largest effect for the
largest grains. To first order in $\Delta v_r/v_k$, the
orbital energy and semimajor axis are unchanged. The range of motion
is therefore centered on the launch distance and has magnitude $2ae =
2 r_L \Delta v_r/v_k \approx 0.067R_p$ for the parameters in
Fig.~\ref{fig:2}b.
 
\begin{figure} [placement h]
\includegraphics [height = 2.1 in] {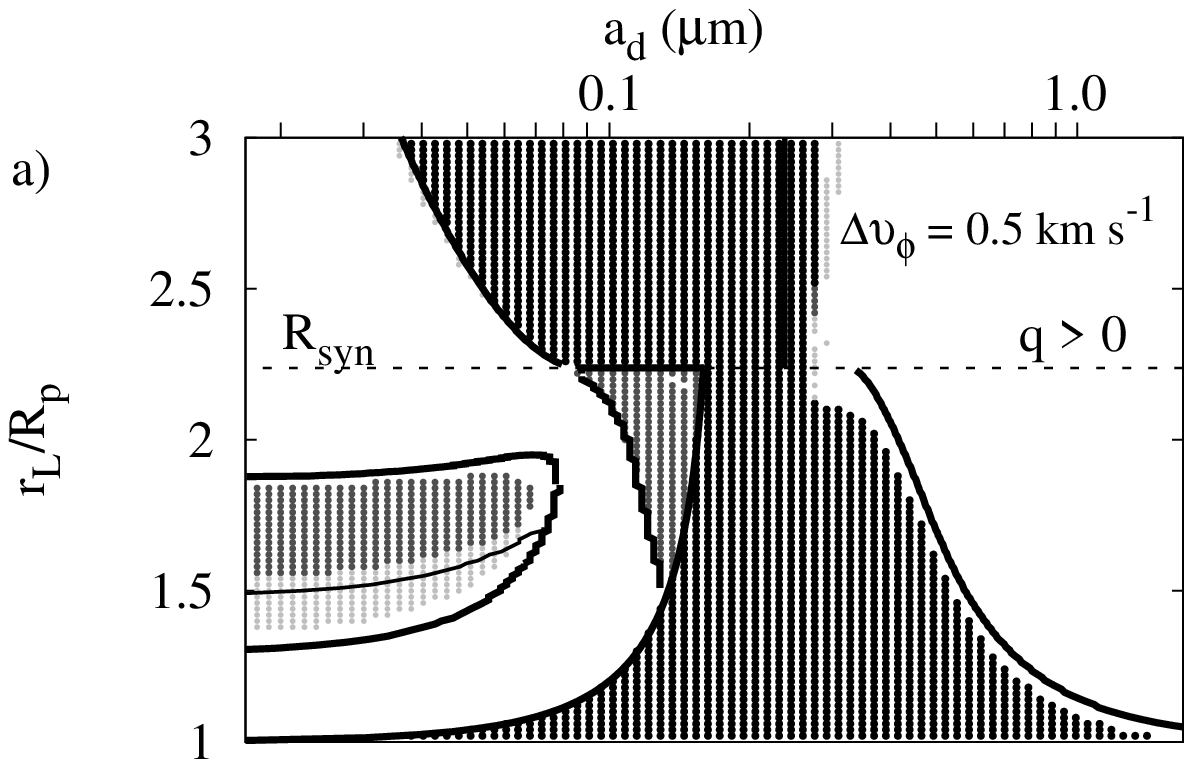}
\newline
\includegraphics [height = 1.689 in, width = 3.0 in] {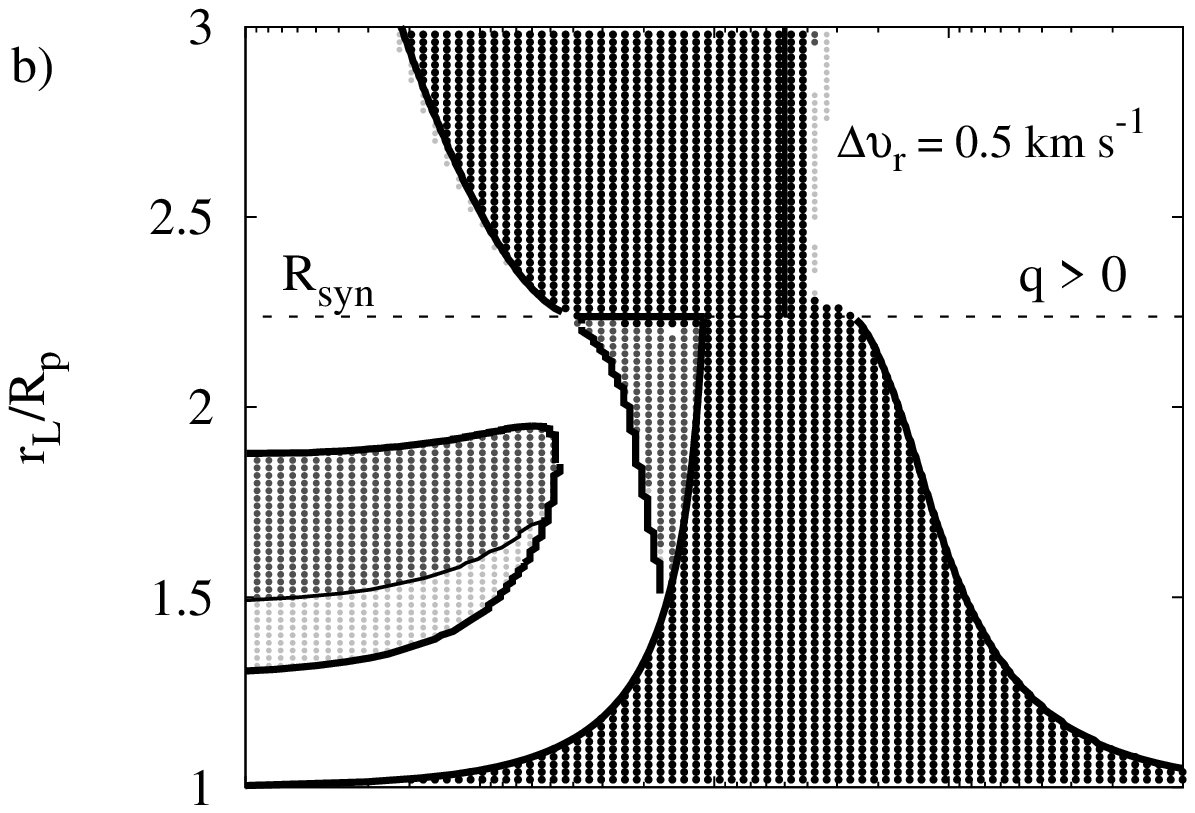}
\newline
\includegraphics [height = 2.1 in] {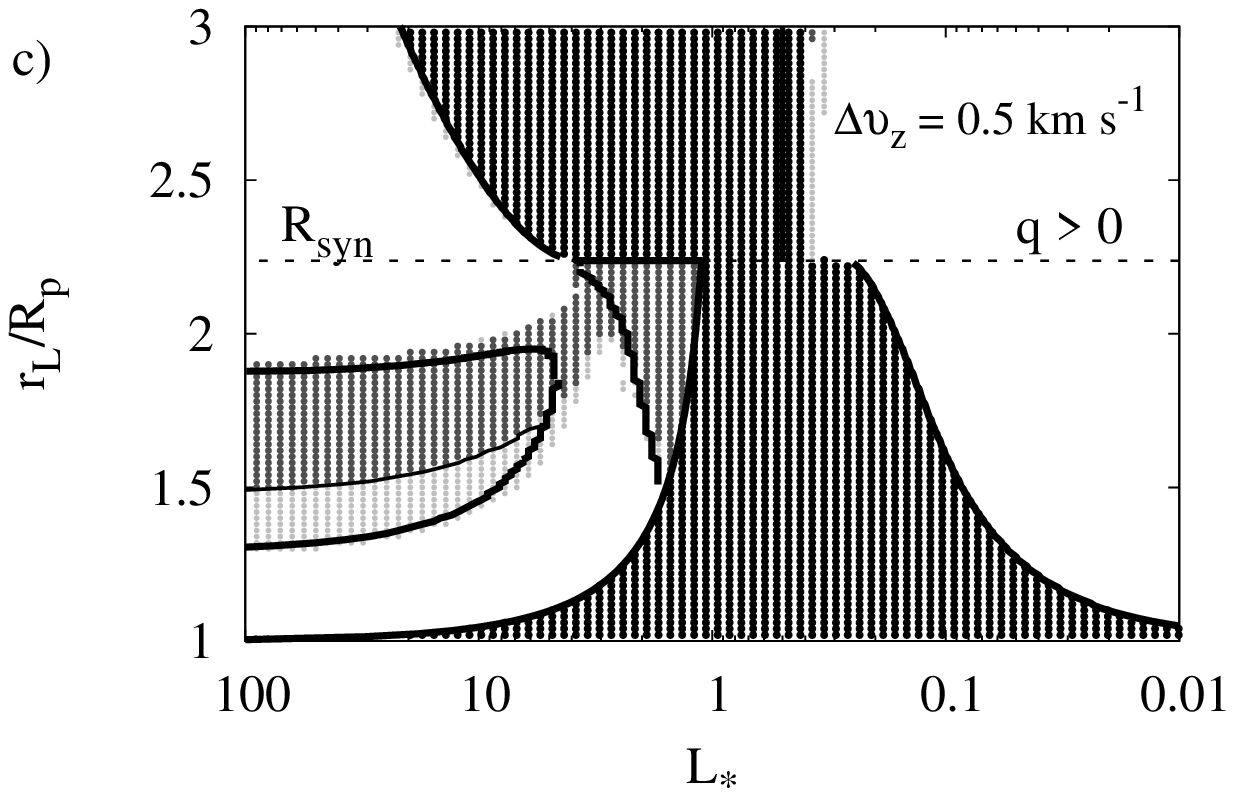}
\caption{Stability of positive grains launched with a) an
  azimuthal launch speed $\Delta v_{\phi} =$ 0.5 km s$^{-1}$ faster
  than the local Kepler speed, b) $\Delta v_{r}$ = 0.5 km s$^{-1}$,
  and c) $\Delta v_{z}$ = 0.5 km s$^{-1}$, integrated over 0.1
  years. The solid curves are the numerically-determined stability boundaries of
  Fig.~\ref{fig:1} where grains are launched at the local circular
  Kepler speed. The darkest regions denote
  grains that collide with the planet near the equatorial plane
  or escape. Moderate grey marks grains that collide with
  the planet at high latitudes ($\lambda > \lambda_m = 5^{\circ}$),
  while light grey shows HLO grains; locally unstable with vertical oscillations exceeding 5$^{\circ}$, but bound globally. White regions mark locally-stable trajectories. Here, radial
  stability (central regions) is only significantly affected by $\Delta v_{\phi}$ while vertical stability (left-most regime) is affected by $\Delta v_{\phi}$ and
  $\Delta v_{z}$. A radial impulse, $\Delta v_r$, has no
  noticeable effects. }
\label{fig:Lmapvvar}
\end{figure}

Figure~\ref{fig:Lmapvvar} highlights the effect of velocity impulses
on the stability boundaries of Fig.~\ref{fig:1} for positively-charged
grains in three orthogonal directions: a prograde azimuthal impulse
($\Delta v_{\phi} = +0.5$ km s$^{-1}$), a radial boost ($\Delta v_r =
+0.5$ km s$^{-1}$), and a vertical kick ($\Delta v_z = +0.5$
km s$^{-1}$). In each case, the orbital stability boundaries are only
moderately affected by these changes; circular orbits are
thus often a good approximation when considering stability. Only the
azimuthal impulse appreciably affects the orbital energy, and hence
shifts the radial stability boundary (larger grains on the right in
Fig.~\ref{fig:Lmapvvar}a). In this case, the positive $\Delta
v_{\phi}$ increases the Kepler orbital energy significantly, thereby
preventing grains near the right-most radial stability boundary from
falling into Jupiter. A negative $\Delta v_{\phi}$ would destabilize
grains near this boundary, permitting additional grains to fall to
the planet. The left side boundary of the radially unstable zone is
basically unaffected by all impulses, in agreement with
Fig.~\ref{fig:2}.

The vertical stability boundaries are moderately affected by the
$\Delta v_{\phi}$ (Fig.~\ref{fig:Lmapvvar}a) and $\Delta v_{z}$
(Fig.~\ref{fig:Lmapvvar}c) initial impulses, but a radial impulse
(Fig.~\ref{fig:Lmapvvar}b) has almost no discernable effect. Note that
although in the Lorentz limit, the radial range of motion is too small
to be significantly altered with a $\Delta v_{\phi}$ launch impulse,
the change in the area of a gyroloop, which alters the magnetic mirror force, still noticeably affects the high
$\Lpar$ stability boundary in Fig.~\ref{fig:Lmapvvar}a. Since $\Delta
v_{\phi} > 0$ in Fig.~\ref{fig:Lmapvvar}a, the increased gyrospeed
expands the gyroloop, leading to a stronger mirror force and hence a
reduced region of vertical instability (Eq.~\ref{Omb}). Enhanced instability results
for $\Delta v_{\phi} < 0$. A vertical impulse $\Delta v_{z}$ of either
sign also leads to additional instability
(Fig.~\ref{fig:Lmapvvar}c). For moderate values of $\Lpar$ in
particular, the $\Delta v_{z}$ impulse causes the vertical instability
region to dramatically expand near $\Lpar = 3, r_L = 2 R_p$, and merge
with the HRLO region of large radial and vertical
oscillations (cf. Fig.~\ref{fig:1}). Negatively-charged grains (not shown)  are
similarly affected by 0.5 km s$^{-1}$ impulses.

Overall, since the majority of real debris particles have much smaller
speeds relative to their parent satellites than the 0.5 km s$^{-1}$ considered here, we conclude that the
stability boundaries are fairly insensitive to grain launch
conditions. We note that new stability boundaries appropriate for
non-circular initial orbits could be derived analytically using
Hamiltonian methods (\citealt{nh82,sb94,mhh03,jh12}), but as the
effect is unimportant for our purposes, we turn instead to more
complicated magnetic field geometries.
\subsection{Vertically Offset Dipole}
In this section we isolate the effect on orbital stability of Jupiter's dipole offset, modelled by the $g_{10}$ and $g_{20}$ magnetic field terms.
\begin{figure} [placement h]
\includegraphics [height = 2.1 in] {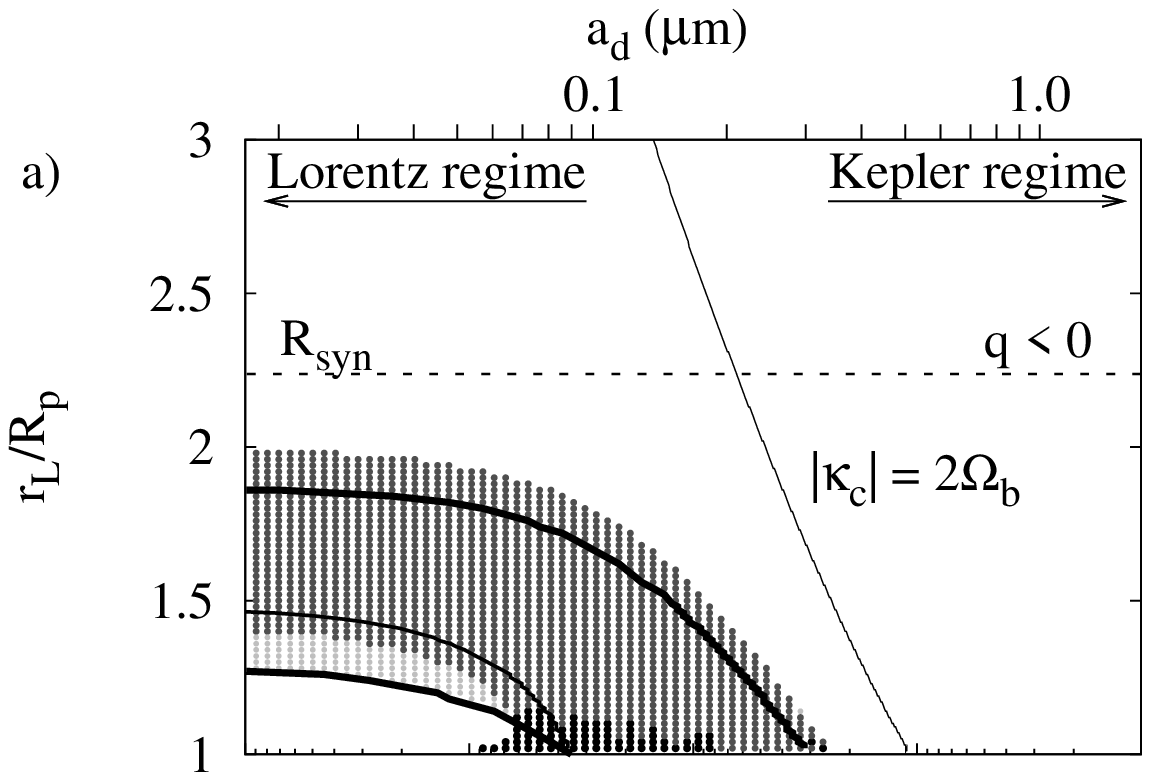}
\newline
\includegraphics [height = 2.1 in] {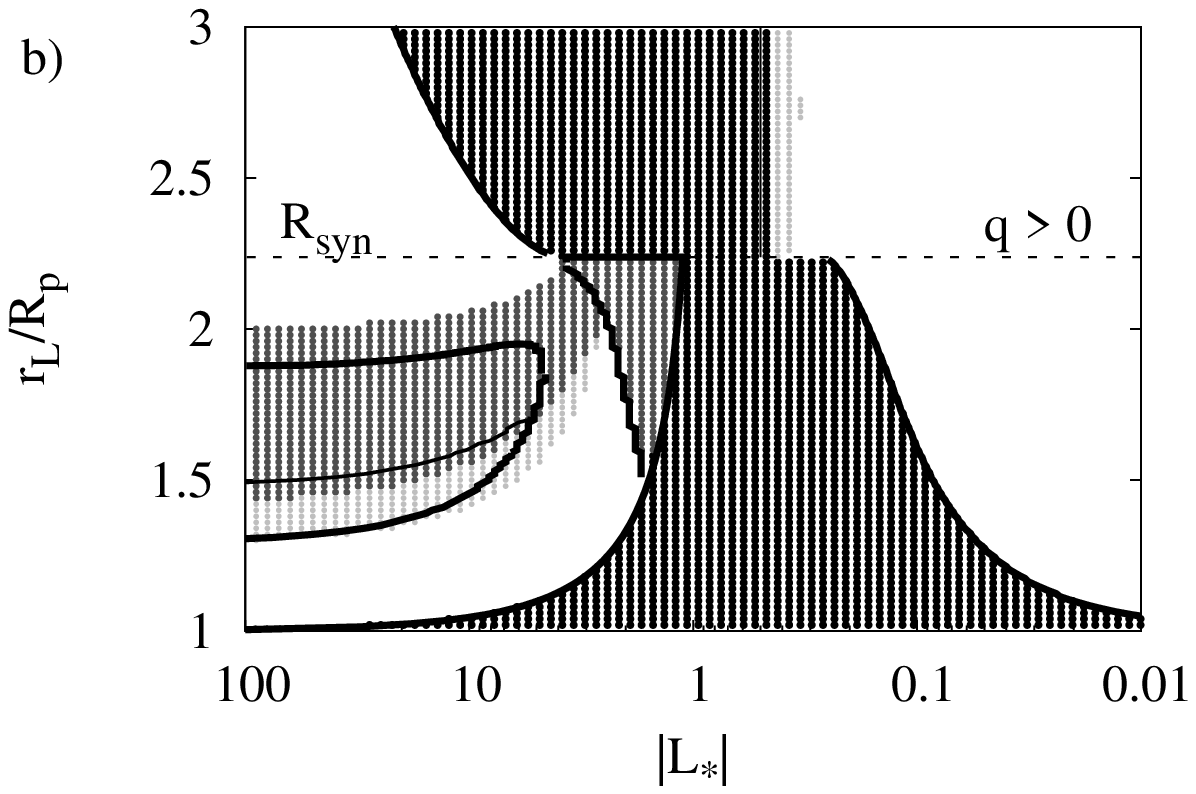}
\caption{Stability of Kepler-launched negative grains (a) and positive
  grains (b) in a \textbf{vertically offset dipole field} for Jupiter,
  modelled with just the $g_{10}$ and $g_{20}$ magnetic field terms,
  and integrated for 0.1 years. The curves indicate the numerical stability
  boundaries for the centered and aligned dipole configurations from
  Fig.~\ref{fig:1}. In this map, the dark areas are radially unstable
  grains that either hit the planet or escape at low latitude
  ($|\lambda| < \lambda_m = 5^{\circ}$). The moderately-grey regions
  are vertically unstable grains that collide with the planet at high
  latitude ($|\lambda| > \lambda_m$), and the light grey regions show
  HLO stable grains that exceed $\lambda_m$ in latitude, the same
  criteria that we have adopted for the aligned dipole magnetic field of
  Fig.~\ref{fig:Lmapvvar}.}
\label{fig:g20negpos}
\end{figure} 
The maps in Fig.~\ref{fig:g20negpos} show that the offset field
exacerbates the vertical instability for both negative and positive
grains but has little effect on the radial stability boundaries. For
the positive grains, the vertically unstable and HRLO zones overlap as
in Fig.~\ref{fig:Lmapvvar}c.

In the equator plane, the $g_{20}$ magnetic field is radial, and the
corresponding $\vec{v}\times \vec{B}$ force is vertical. Thus the
$g_{20}$ magnetic field term primarily adds an additional vertical force, thereby
expanding the vertical instability. In fact, the stability map in
Fig.~\ref{fig:g20negpos}b resembles Fig.~\ref{fig:Lmapvvar}c, which
modelled a vertical impulse on the grains at launch; comparison of the
two figures shows that the inclusion of $g_{20}$ is a far more important effect. Note
also that, unlike the effect of a $\Delta v_z$ impulse, the offset
dipole is effective at destabilizing grains near the planet, causing
a significant vertical bounce oscillation and significantly expanding the region of global vertical
instability.

\subsection{Tilted Dipole}
\begin{figure} [placement h]
\includegraphics [height = 2.1 in] {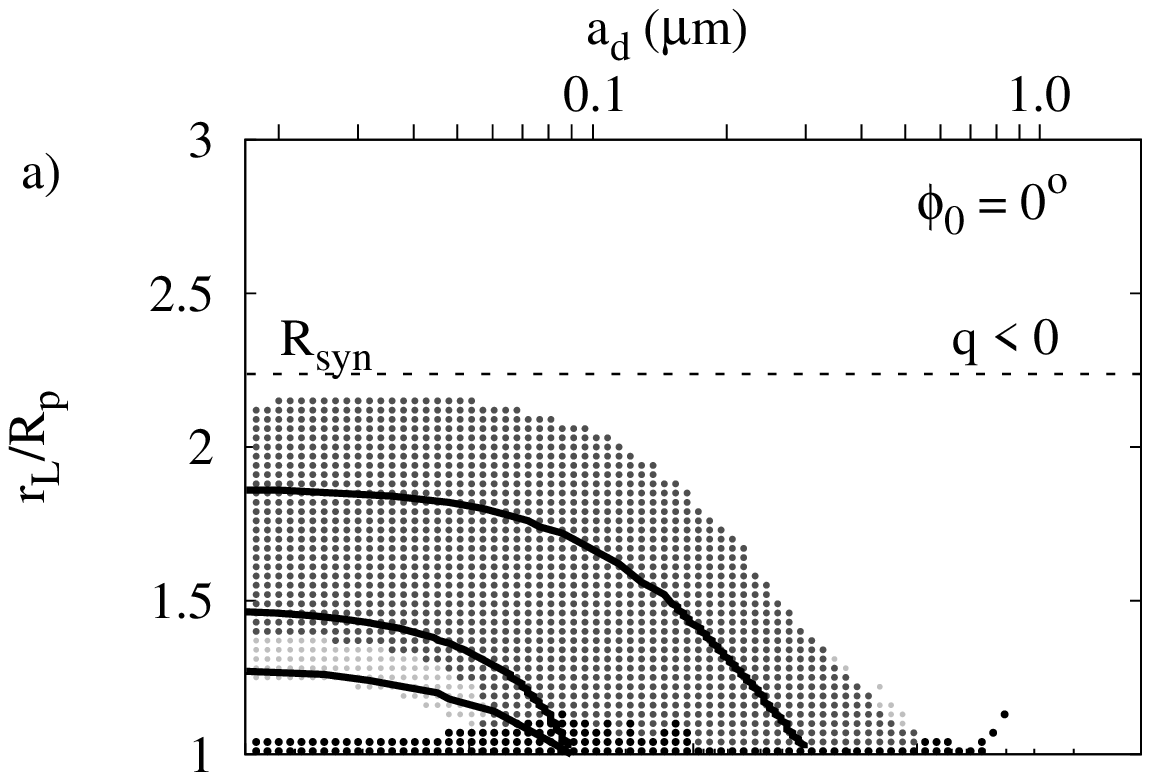}
\newline
\includegraphics [height = 2.1 in] {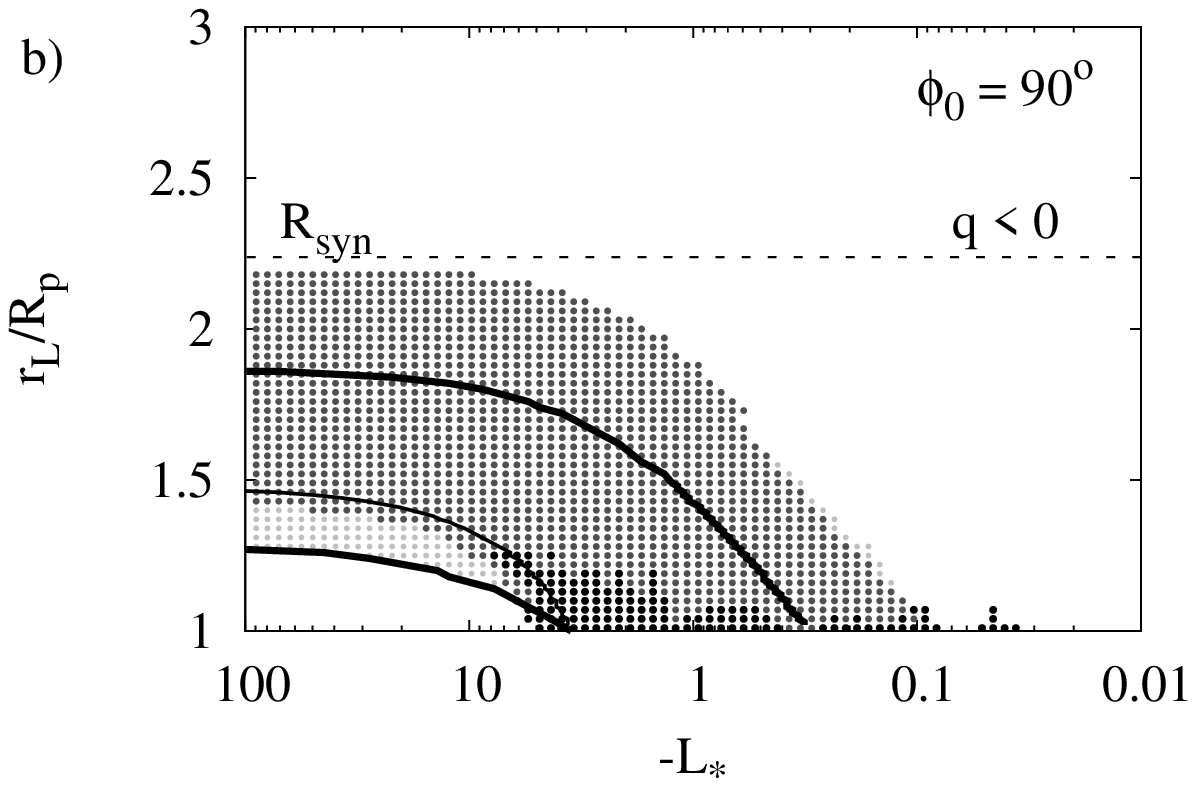}
\caption{Stability of negative grains integrated over 0.1 years in a
  \textbf{tilted dipole field} for Jupiter, with the launch longitude
  at two locations; a) where the magnetic equator crosses the
  planetary equator plane ($\phi_0 =0^{\circ}$), and b) where the
  magnetic equator reaches its highest geographic latitude
  ($\phi_0=90^{\circ}$). As before, the curves are the numerically-determined stability boundaries for the aligned dipole case from
  Fig.~\ref{fig:1}a. The grey scale is similar to
  Figs.~\ref{fig:Lmapvvar} and~\ref{fig:g20negpos}: dark points are
  grains that collide with the planet at low latitudes, the moderately-grey region denotes grains that were vertically unstable to collide
  with the planet at latitudes higher than $\lambda_m = 20^{\circ}$,
  the lightest grey marks trajectories that were excited to higher
  latitudes but remained bound, and the white areas represent grains
  that were locally stable. The only difference from
  Fig.~\ref{fig:Lmapvvar} and~\ref{fig:g20negpos} is that here we
  define low latitude to be $|\lambda_m| < 20^{\circ} $ rather
  than 5$^{\circ}$.}
\label{fig:g11fineg}
\end{figure}

In testing the effect of a tilted dipole field, we include the
$g_{10}$, $g_{11}$ and $h_{11}$ magnetic field terms in our numerical models, setting
$g_{20}$ and all higher order terms to zero. Since the magnetic and
gravitational equators do not coincide, we consider two separate
equatorial launch phases: (i) $\phi_{0} = 0^{\circ}$, the ascending node of
the magnetic equator on the geographic equator and (ii) $\phi_0 =
90^{\circ}$, where the magnetic equator reaches its highest northern
latitude of 9.6$^{\circ}$. At this launch phase, many grains can reach
latitudes $\approx 20^{\circ}$ north and south of the equator, even if
their trajectories are stable. Our stability results for
negatively-charged grains are plotted in Fig.~\ref{fig:g11fineg}.

Although slight differences with launch phase are apparent,
Figs.~\ref{fig:g11fineg}a and~\ref{fig:g11fineg}b are quite
similar. Jupiter's tilt is a stronger effect than its offset (see
Fig.~\ref{fig:g20negpos}a), extending the vertical instability
boundary significantly outwards and close to $R_{syn}$. The dramatic
outward expansion of the vertical instability can be understood as
follows. For an aligned dipole, as synchronous orbit is approached,
both the velocity relative to the magnetic field and the
electromagnetic forces tend toward zero. Furthermore, as the velocity is azimuthal and the field is vertical, the direction of the weak EM force is entirely radial. For a tilted dipole,
however, the magnetic field lines cross the equator plane with a
radial component, causing a substantial $\vec{v} \times \vec{B} $
vertical force as with the offset dipole. These forces push particles
out of the plane along field lines, leading to an expansion of the instability zone nearly to $R_{syn}$. Interestingly, the inner boundary at ($\Lpar = -50$,
$r_L/R_p = 1.4$) is far less affected.

One key difference in the two panels of Fig.~\ref{fig:g11fineg} occurs
for high $\Lpar$ along $r_L/R_p = 1$; launching at the node ($\phi_0 =
0^{\circ}$) leads to collisions while launching at $\phi_0 =
90^{\circ}$ does not. This difference is due to the curvature of the
field lines in a dipole. In an aligned dipole magnetic field, stable
mirror motion causes EM-dominated grains to oscillate about the
magnetic equator, whereby the turning points or mirror points confine
the latitudinal range of the grain. Launching from $\phi_0 =
90^{\circ}$ in the tilted magnetic field ensures that the launch point
is at one of the mirror points, and this vertical turning point is
relatively close to Jupiter. Thus grains launched near 1$R_p$ in Fig.~\ref{fig:g11fineg}b 
initially move radially outward and do not collide with the planet. By
contrast, for grains launched at the node where $\phi_0 = 0^{\circ}$,
the mirror points are necessarily closer to the planet than the launch
distance and, accordingly, we see that grains launched within
$1.06R_p$ are forced to collide with Jupiter in Fig.~\ref{fig:g11fineg}a.

Another, more subtle difference between Figs.~\ref{fig:g11fineg}a
and~\ref{fig:g11fineg}b, is that, grains launched
at the node ($\phi_0 = 0^{\circ}$) are slightly more stable close to
$R_{syn}$ than those launched at $\phi_0 = 90^{\circ}$ e.g. at ($\Lpar
= -100$, $r_L/R_p = 2.21$). In fact, both the inner and outer vertical
stability boundaries are shifted slightly
outwards for $\phi_0 = 90^{\circ}$ compared to launches at $\phi_0 =
0^{\circ}$.
\begin{figure} [placement h]
\includegraphics [height = 2.1 in] {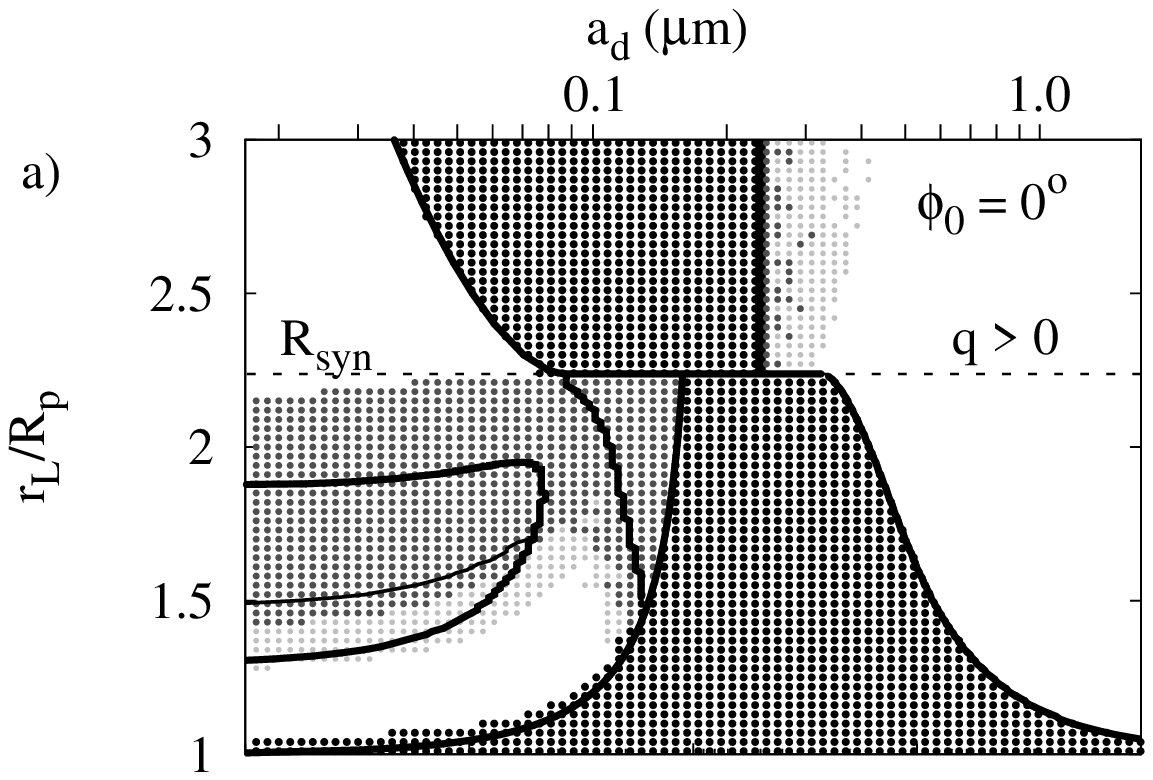}
\newline
\includegraphics [height = 2.1 in] {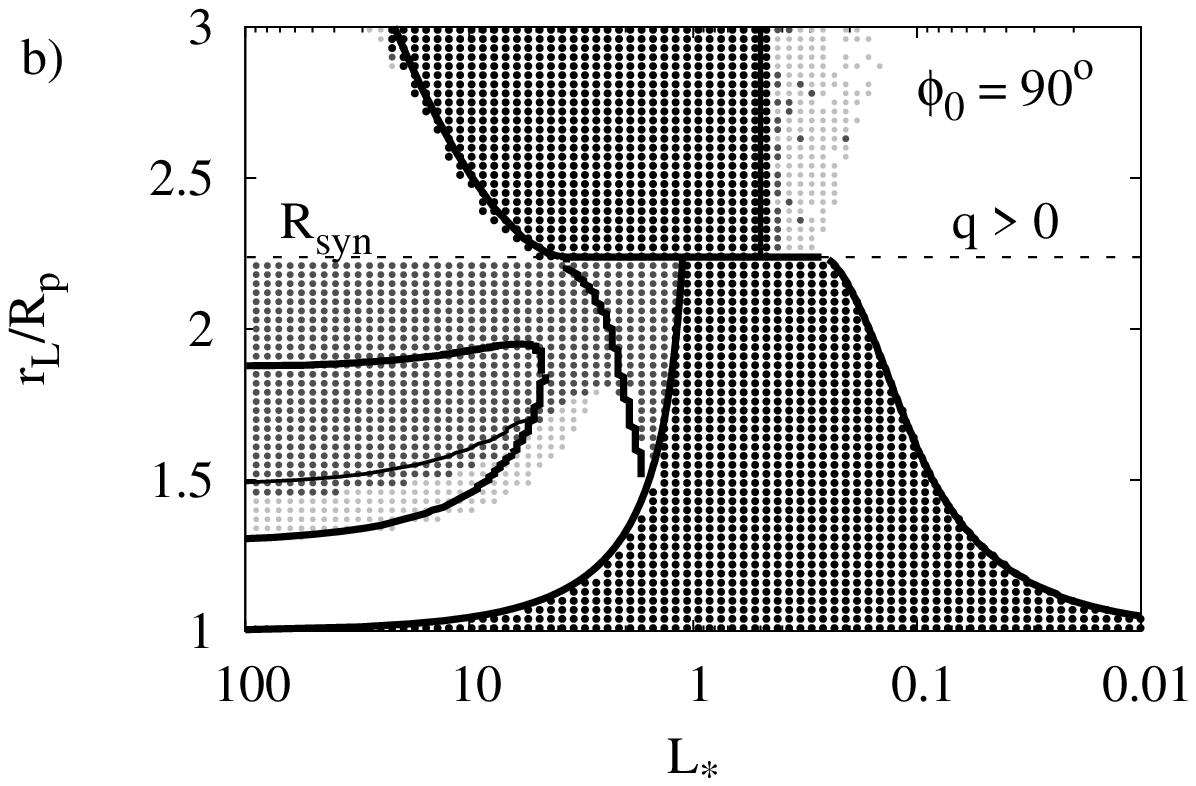}
\caption{Stability of positive grains integrated for 0.1 years in a
  \textbf{tilted dipole field} for Jupiter, with two different launch
  longitudes: a) $\phi_{0} = 0^{\circ}$, where the magnetic equator
  crosses the planetary equator plane, and b) $\phi_0 = 90^{\circ}$,
  where the magnetic equator reaches its highest geographic
  latitude. The solid black curves mark the stability boundaries for
  the aligned dipole case from Fig.~\ref{fig:1}b. As in
  Fig.~\ref{fig:g11fineg}, the darkest points denote grains that
  strike the planet at low latitude ($|\lambda_m| < 20^{\circ}$,
  within twice the tilt angle), the moderate grey marks grains that
  strike the planet at high latitudes, the lightest grey marks grains
  that remain bound between high-latitude mirror points, and the white
  area represents grains that are vertically stable.}
\label{fig:g11fipos}
\end{figure}

Positively-charged dust grains are similarly affected by the addition
of the dipole tilt. As with negative grains, we present two launch
phases in Fig.~\ref{fig:g11fipos}, and find differences in orbital
stability similar to those already discussed for
Fig.~\ref{fig:g11fineg}. In particular, the azimuthal dependencies for
highly-charged dust grains ($|\Lpar| >> 1$) are almost identical for
both negative and positive charges (Fig.~\ref{fig:g11fipos}). As in Fig.~\ref{fig:g11fineg}a, grains launched immediately above the planet are unstable for $\phi_0 = 0^{\circ}$ (Fig.~\ref{fig:g11fipos}a).

For all launch longitudes, the vertical instability expands greatly outwards,
nearly to synchronous orbit but very little towards the planet. As in Fig.~\ref{fig:g11fineg}, both the inner and outer vertical
stability boundaries in the Lorentz limit are shifted slightly
outwards for the $\phi_0 = 90^{\circ}$ launches of
Fig.~\ref{fig:g11fipos}b compared to the $\phi_0 = 0^{\circ}$ launches
of Fig.~\ref{fig:g11fipos}a. The two boundaries have slightly
different explanations. For the outer boundary near $R_{syn}$,
launching at $\phi_0 = 90^{\circ}$ allows the initially larger
vertical electromagnetic forces to drive the grain to higher latitudes
where the gravity of the planet can overwhelm the centrifugal force
and cause instability \citep{jh12}. Near the inner vertical boundary, however,
grains launched at the node $\phi_0 = 0^{\circ}$ have a higher
latitudinal range and are slightly less stable.

The radial stability boundaries are largely unaffected by the tilt in
the magnetic field, although some slight differences are evident to
the left of the radial instability region. Note the subtle difference
along the left-most radial stability boundaries between
Fig.~\ref{fig:g11fipos}a and Fig.~\ref{fig:g11fipos}b, where grains
launched at $\phi_0 = 90^{\circ}$ inside $R_{syn}$, are slightly more
stable than those with $\phi_{0} = 0^{\circ}$. Outside synchronous
orbit, however, the reverse holds true. This is most easily understood
as an overall outward shift of the instability region from $\phi_0 =
90^{\circ}$ to $\phi_0 = 0^{\circ}$. Thus $\phi_0 = 90^{\circ}$ grains
behave almost exactly like $\phi_0 = 0^{\circ}$ grains that have been
launched a bit further from the planet. Note that this difference with
launch phase was also seen for the negative grains with $r_L \approx R_p$
(Fig.~\ref{fig:g11fineg}), and the explanation is the same.

Until now we have only considered instabilities that remove a grain
typically within a few hours. However, the tilted dipolar field causes
further instabilities acting over weeks to months, and over a greater
range of launch distances than the aligned dipolar case. We explore these longer-term effects below.
\subsection{Resonant Effects in a Tilted Dipole Field}
In an aligned dipole field, it can be shown that negative grains
outside $R_{syn}$ are permanently confined between their launch
distance and $R_{syn}$; they are energetically unable to escape
\citep{jh12}. The tilted dipole field however, permits radial motion
away from $R_{syn}$, and actually enables some negative grains to
depart the planet, as was first seen by \citet{ham96}.
\begin{figure} [placement h]
\includegraphics [height = 2.1 in] {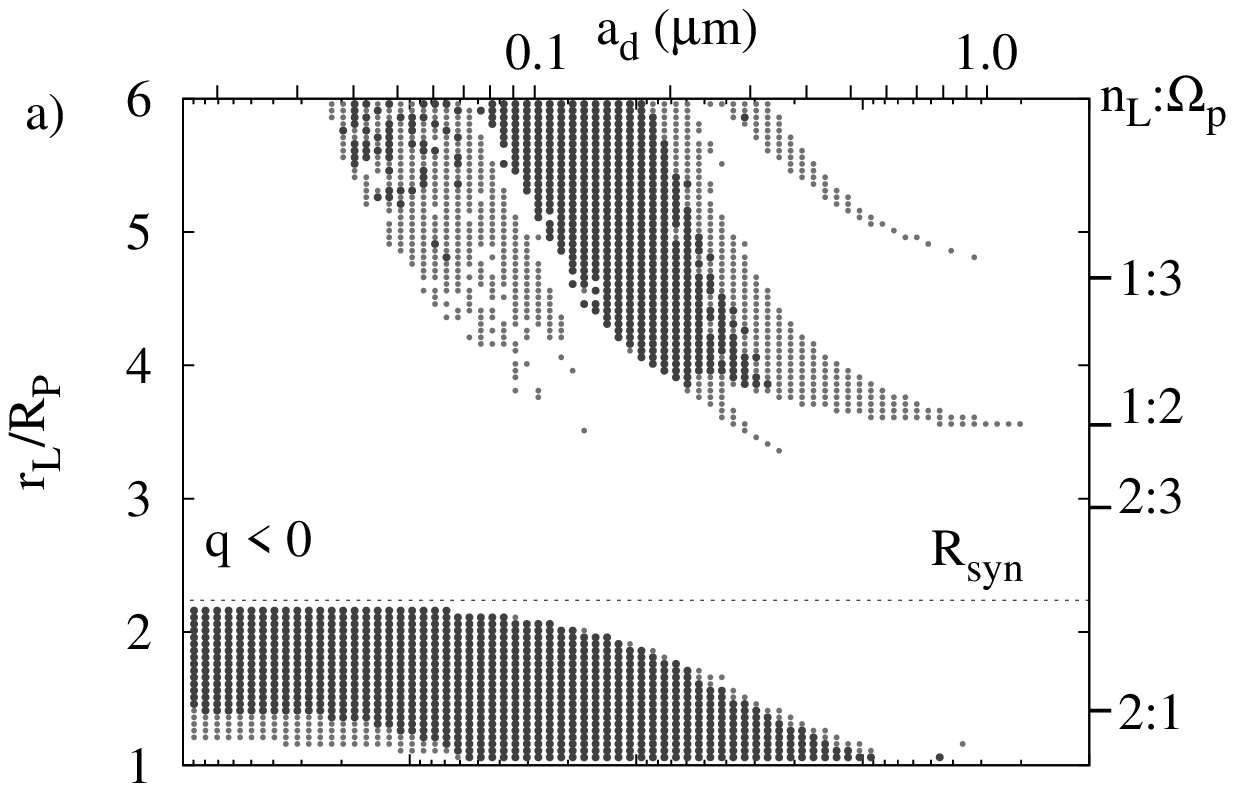}
\newline
\includegraphics [height = 2.1 in] {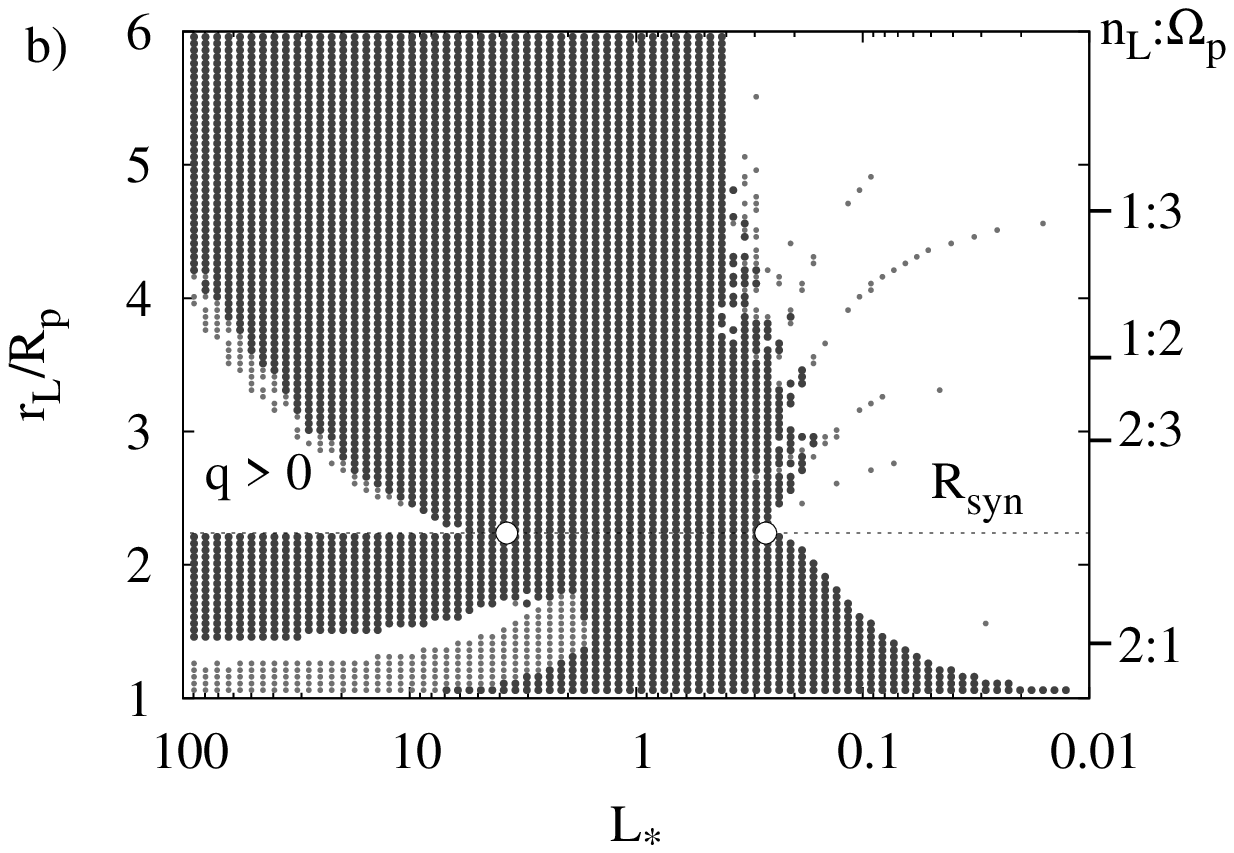}
\caption{Destabilising effects of a \textbf{tilted dipole} magnetic
  field for Jupiter, with launch at $\phi_0 =
  90^{\circ}$. Integrations are for one Earth year, and we show a
  greater radial range than in Figs.~\ref{fig:g11fineg} and
  ~\ref{fig:g11fipos}. Note that the greyscale that we use here is
  also different from the earlier figures. Here, the darkest region denotes
  grains that either escape or crash into Jupiter within 1 year of
  launch. For the negative grains, the light grey marks grains with
  radial motions \textit{away} from $R_{syn}$ (in the direction
  opposite that expected for gyromotion), by at least 0.04$r_L$,
  revealing the destabilizing effect of the tilted magnetic field. For
  positive grains, the light grey indicates trajectories with radial
  motions towards $R_{syn}$ of at least 0.02$r_L$. As always, white indicates stability.}
\label{fig:g10g11}
\end{figure} 

In Fig.~\ref{fig:g10g11}, we show the stability maps for Jupiter
modelled with the $g_{10}$ and $g_{11}$ magnetic field components for
dust grains with both negative and positive charges whose trajectories
were integrated for one Earth year. We seek to highlight the motion
away from synchronous orbit for the negative grains, and towards
$R_{syn}$ for the positive grains, motions precluded by a simple
aligned dipolar magnetic field.

Within synchronous orbit, the negatively-charged grains of
Fig.~\ref{fig:g10g11}a shows the same short-term instabilities seen in
Fig.~\ref{fig:g11fineg}b. Notice in Fig.~\ref{fig:g10g11}a,
however, the large fingers of instability outside synchronous
orbit. These features trace grains that suffer significant and unusual
motions away from $R_{syn}$, and the largest one points towards $r_L/R_p = 3.55$
which happens to be the location of the outer 1:2 Lorentz resonance
(\citealt{sb87,sb92,ham94,sho08}). The other fingers point towards other Lorentz resonances. Importantly, the highly-detailed
dark grey structures within these fingers (Fig.~\ref{fig:g10g11}a) indicate significant numbers of negative
grains that actually escape from Jupiter within one year.

These results are important for the escape of dust from the Io plasma torus, the
most likely source of the jovian high-speed dust streams
(\citealt{hmg93a,gra00}). Dust streams are comprised of radially-accelerated positively-charged dust grains, although in the plasma
torus itself, dust-grain electric potentials are likely to be
negative, even in sunlight (\citealt{bag94,kru03}). Lorentz resonances
can provide a rapid escape mechanism for negatively-charged grains
launched in the plasma torus. Once dust grains are free of the torus,
charging currents become positive and the grains are accelerated outwards to
escape. Our modelling in Fig.~\ref{fig:g10g11}a actually understates the importance of this mechanism, as we force the potential to remain negative far outside the actual boundaries of the plasma torus.

For the positive grains, Fig.~\ref{fig:g10g11}b shows an increased
number of grains that are unstable, compared to
Fig.~\ref{fig:g11fipos}b. The most striking difference is that in
Fig.~\ref{fig:g10g11}b, there are rough patches of additional radial
instability just outside $R_{syn}$ near ($\Lpar \approx 0.2$, $r_L/R_p
= 3$). These unstable patches transition smoothly to become thin
tracks of bound grains with excited radial ranges in the Kepler regime
which, like the negative grains in Fig.~\ref{fig:g10g11}a, point
towards Lorentz resonances which occur for discrete integer ratios of the planetary spin and Kepler
orbital frequencies.

Accordingly, we look to extend the concept of Lorentz resonances,
(much studied in the Kepler limit by authors including
\citealt{bur85,sb87,sb92,hb93b,ham94}), to cover the entire range of
charge-to-mass ratios. To determine the location of these Lorentz
resonances in general, we start with the resonant equation
\begin{equation}
\dot{\Psi} =  A \omega_c+B\Omega_p+C\dot{\Omega}_{node} + D\dot{\varpi}_{peri},
\label{Lres1}
\end{equation}
where the coefficients A, B, C and D are integers that must sum to
zero \citep{ham94}. Here $\omega_c$ is the orbital frequency of the
guiding center, $\Omega_p$ is the planetary spin rate,
$\dot{\Omega}_{node}$ is the precession rate of the ascending node,
$\dot{\varpi}_{peri}$ is the precession rate of the pericenter, and
$\Psi$ is the resonant argument; $\dot{\Psi}$ equals zero at a Lorentz
resonance.

Equation~\ref{Lres1} is completely general and valid for all charge-to-mass ratios if we are careful to rewrite $\dot{\varpi}_{peri}$ and $\dot{\Omega}_{node}$ in
terms of our fully general frequencies from Eqs.~\ref{omegac} -~\ref{Omb}. The precession rates are simply differences between fundamental frequencies:
\begin{equation}
\dot{\varpi}_{peri} = \omega_c-|\kappa_c| 
\label{pericenter}
\end{equation}
and
\begin{equation}
\dot{\Omega}_{node} = \omega_c-\Omega_b 
\label{ascendnode}
\end{equation}
where $\kappa_c$, the epicyclic frequency of motion, is negative by
convention for retrograde epicycles. Recall that in the gravity
limit, $\omega_c = |\kappa_c| = \Omega_b = n_c$ and hence
$\dot{\varpi}_{peri} = \dot{\Omega}_{node} = 0$, as expected. As an
illustration, we focus on radial resonances for which $C = 0$.

Setting Eq.~\ref{Lres1} to zero and using Eq.~\ref{pericenter} to eliminate $\dot{\varpi}_{peri}$, we find:
\begin{equation}
-B\omega_c + B\Omega_p -D|\kappa_c| = 0
\label{Lres2}
\end{equation}
or
\begin{equation}
B\dot{\phi_c}+D|\kappa_c| = 0,
\label{Lres3}
\end{equation}
since in the frame corotating with the magnetic field, the azimuthal
frequency of the guiding center is given by $\dot{\phi_c} = \omega_c -
\Omega_p$.

Equation~\ref{Lres3} shows that a Lorentz resonance affecting radial
oscillations reduces to a simple ratio between the epicyclic frequency
$|\kappa_c|$, and the motion of the guiding center relative to the
rotating magnetic field ($\dot{\phi_c}$). Our approach thus shows how to extend classical Lorentz resonances to remain valid at arbitrary charge-to-mass
ratios.
\begin{table}
\begin{center}
\begin{tabular}{||l|l|l|l|l||}
\hline
 $\vec{B}$-field terms &  Res. name & $r_L$ ($\Lpar \rightarrow 0$) & Res. frequency $\dot{\Psi}$ & Corotating form \\
  \hline
  \hline
   $g_{11}$  &  1:4 & 5.64  $R_p$  & $4\omega_c-\Omega_p-2\dot{\varpi}_{peri}-\dot{\Omega}_{node}  $ & $\dot{\phi}_c+2|\kappa_c| + \Omega_b$  \\
  \hline
  $g_{11}$  &   1:3 & 4.66  $R_p$  & $3\omega_c-\Omega_p-\dot{\varpi}_{peri}-\dot{\Omega}_{node}  $ & $\dot{\phi}_c+|\kappa_c| +\Omega_b $  \\
  \hline
   $g_{11}$  &  1:2 & 3.55  $R_p$  & $2\omega_c-\Omega_p-\dot{\Omega}_{node}  $ & $\dot{\phi}_c+ \Omega_b$ \\
  \hline
\hline
  $g_{21}$   & 1:3 &  4.66  $R_p$ & $3\omega_c-\Omega_p-2\dot{\varpi}_{peri}  $ & $\dot{\phi}_c+2|\kappa_c|$  \\
  \hline
  $g_{21}$  &   1:3 & 4.66  $R_p$  & $3\omega_c-\Omega_p-2\dot{\Omega}_{node}  $ & $\dot{\phi}_c+2\Omega_b$  \\
  \hline
  $g_{21}$  &  1:2 & 3.55  $R_p$  & $2\omega_c-\Omega_p-\dot{\varpi}_{peri}  $  & $\dot{\phi}_c+|\kappa_c|$    \\
  \hline
\hline
  $g_{22}$  &   2:4 & 3.55  $R_p$  &$4\omega_c-2\Omega_p-\dot{\varpi}_{peri}-\dot{\Omega}_{node}  $ & $2\dot{\phi}_c+|\kappa_c|+\Omega_b$  \\
  \hline
   $g_{22}$  & 2:3 & 2.93  $R_p$  & $3\omega_c-2\Omega_p-\dot{\Omega}_{node}  $ & $2\dot{\phi}_c+\Omega_b$  \\
  \hline
  $g_{22}$  &   2:1 & 1.41  $R_p$  &$\omega_c-2\Omega_p+\dot{\Omega}_{node}  $ & $2\dot{\phi}_c - \Omega_b$  \\
  \hline
\hline
\end{tabular}
\caption{Selected Lorentz resonances (column 2), driven by the magnetic field
  coefficient in column 1, appear, for small charge-to-mass ratios, at
  the locations given in column 3. The resonance frequency is given in
  its most general form (column 5) and in a second form most useful
  when gravity dominates (column 4).}
\end{center}
\end{table}

In Table 1, we show select Lorentz resonances for all charge-to-mass
ratios and their driving magnetic field terms, taken from
\citet{ham94}. In the Kepler limit, these Lorentz resonances act to
slowly increase eccentricities and/or inclinations, destabilizing
trajectories over many orbits. The resonances that include multiple instances of $\dot{\varpi}_{peri}$ or $\dot{\Omega}_{node}$, such as the three 1:3 and 1:4 resonances in Table 1, are weaker since their strengths in the Kepler regime depend on higher powers of the small quantities $e$ (eccentricity) and $i$ (inclination). At higher charge-to-mass ratios, however, all of these resonances increase in strength, and their effects on grain orbits occur on much shorter timescales than
in the Kepler regime. Some negative grains at 6.0$R_p$ in
Fig.~\ref{fig:g10g11} escape in as little as a few days.

\begin{figure} [placement h]
\includegraphics [height = 2.1 in] {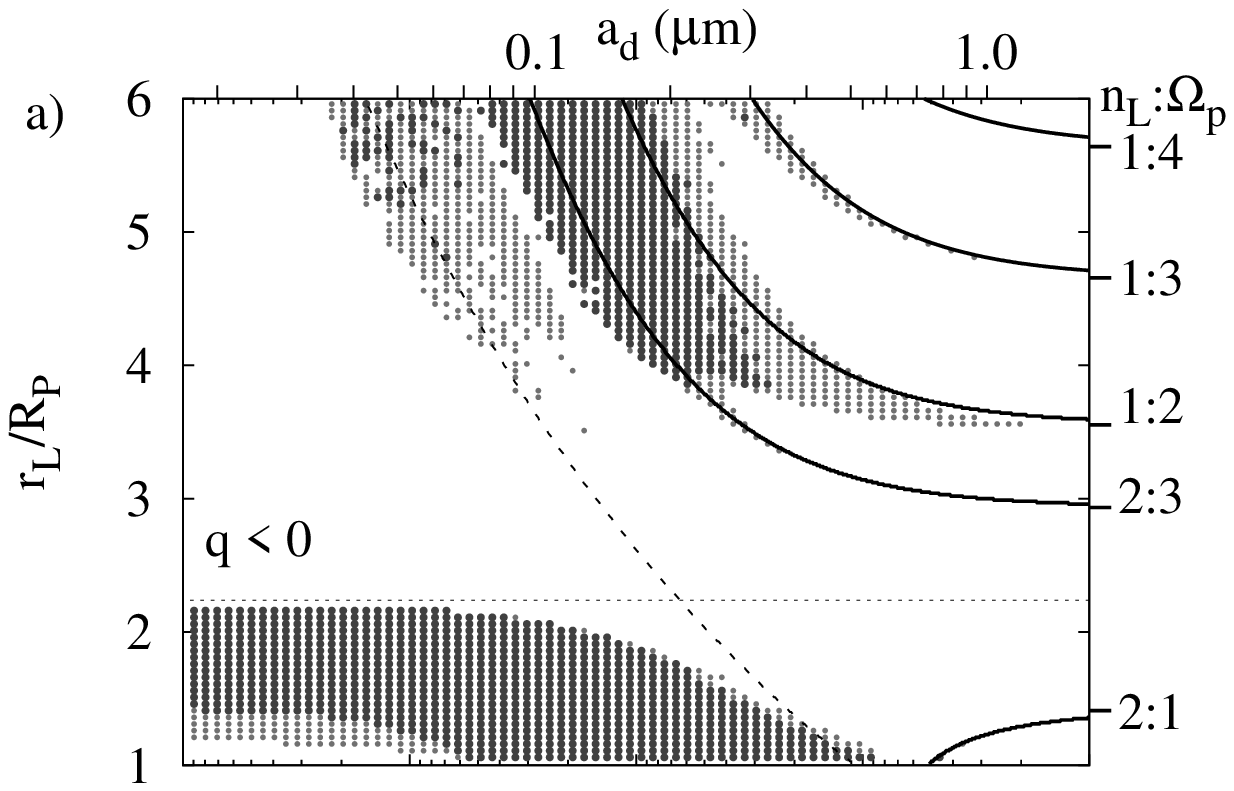}
\newline
\includegraphics [height = 2.1 in] {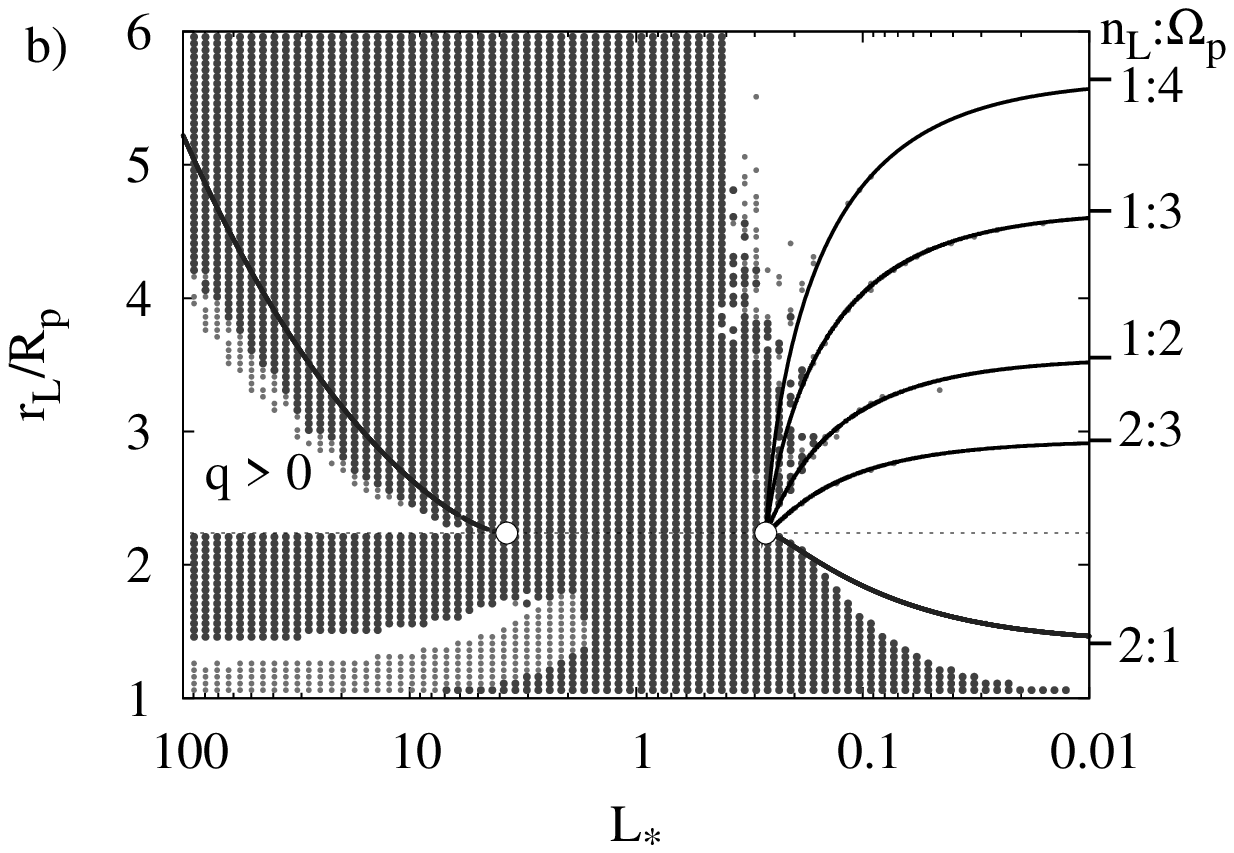}
\caption{The \textbf{tilted dipole} stability map of
  Fig.~\ref{fig:g10g11} with theoretical curves $\dot{\Psi} = 0$ for radial resonances only from
  Table~1 superimposed as solid curves. The theoretical curves fall
  atop the instability ``fingers'' seen in Fig.~\ref{fig:g10g11}a, and the trails of points from Fig.~\ref{fig:g10g11}b,
  attesting to the accuracy of the theory. The 2:1 resonance between
  epicyclic frequency $|\kappa_c|$ and vertical $\Omega_b$ motions is
  also shown in the upper panel, as a dashed curve. In the lower
  panel, the open circles mark the points ($\Lpar = 2\pm\sqrt{3}$,
  $r_L/R_{syn} = 1$), where $\dot{\phi}_c = \kappa_c = 0$ and all resonant tracks for positive
  grains converge. For large $\Lpar$, all radial resonances lie nearly atop one another in panel b). }
\label{fig:g11res}
\end{figure} 

Figure~\ref{fig:g11res} overlays the strictly radial Lorentz resonances of
Table 1 on the stability map for a tilted dipolar field (the data from
Fig.~\ref{fig:g10g11}a). For negative grains outside synchronous orbit,
the Lorentz resonances curve upwards directly into the region of
escape for increasing $\Lpar$. This occurs because the epicyclic
frequency $|\kappa_c|$ increases rapidly with $\Lpar$ (Eq.~\ref{kappa});
$\dot{\phi}_c$ must also increase to maintain a given resonance
(Eq.~\ref{Lres3}). Since $\dot{\phi}_c$ increases away from $R_{syn}$, remaining in resonance as $|\kappa_c|$ increases necessitates a greater launch distance from synchronous orbit. Although these curves are determined from frequencies that
are strictly valid only for an aligned dipole field, they nevertheless show
an impressive match to our data, despite the more complex magnetic
field.

For the positive grains, the resonant tracks in Fig.~\ref{fig:g11res}b
begin at the same locations in the Kepler limit as for negative grains, but they curve towards synchronous orbit as $\Lpar$ increases. On the right side of the short-term radial instability of
Fig.~\ref{fig:1}, all resonant solutions converge to a single
point at ($\Lpar = 2-\sqrt{3}$, $r_L = R_{syn}$). This is the point at
synchronous orbit where $|\kappa_c| = \dot{\phi}_c = 0$, and grain
orbits are locally unstable in even a simple aligned dipolar field
\citep{jh12}. The convergence to synchronous orbit as $\Lpar$
increases occurs because for positive grains with $\Lpar << 1$,
$|\kappa_c|$ decreases as $\Lpar$ increases (Eq.~\ref{kappa}), and so $\dot{\phi}_c$ must decrease as well (Eq.~\ref{Lres3}), driving
distances towards $R_{syn}$.

The Lorentz resonances destabilize the motion of grains, and hint that
a non-axisymetric field allows the negative grains to tap into
planetary rotation, to make escape energetically favorable. The
detailed structure in the stability map of Fig.~\ref{fig:g11res},
including the escaping negatively-charged grains, is due only to the
effects of $g_{10}$ and $g_{11}$. The first-order theory, however, can only
explain the three 1:N resonances (Table 1) and not the instability of the
2:3 and 2:1 resonances, which nevertheless are definitely present in Figs.~\ref{fig:g10g11}
and~\ref{fig:g11res}. We will return to explain this discrepancy
shortly.

In addition to Lorentz resonances of the type shown in
Eq.~\ref{Lres1}, there are also resonances between the dust grain's
radial and vertical motions, analogous to the Kozai resonance
experienced by highly-inclined orbits. The dominant resonance of this
type satisfies: $ \omega_c -2\dot{\Omega}_{node}+\dot{\varpi}_{peri} =
0$, such that $|\kappa_c| = 2\Omega_b$. This 2:1 resonance between
radial and bounce motions is the strongest of its type since during
one bounce period, north-south symmetry ensures that the dust grain
experiences two cycles in magnetic field strength \citep{jh12}. The
resonance track also passes close to the high charge-to-mass boundary
of the resonant structure in Fig.~\ref{fig:g11res}a. We turn now to
investigate the effects of the individual asymmetric quadrupolar
magnetic field terms which should also power resonances (Table 1).

\subsection{Quadrupole Terms}
\begin{figure} [placement h]
\includegraphics [height = 2.1 in] {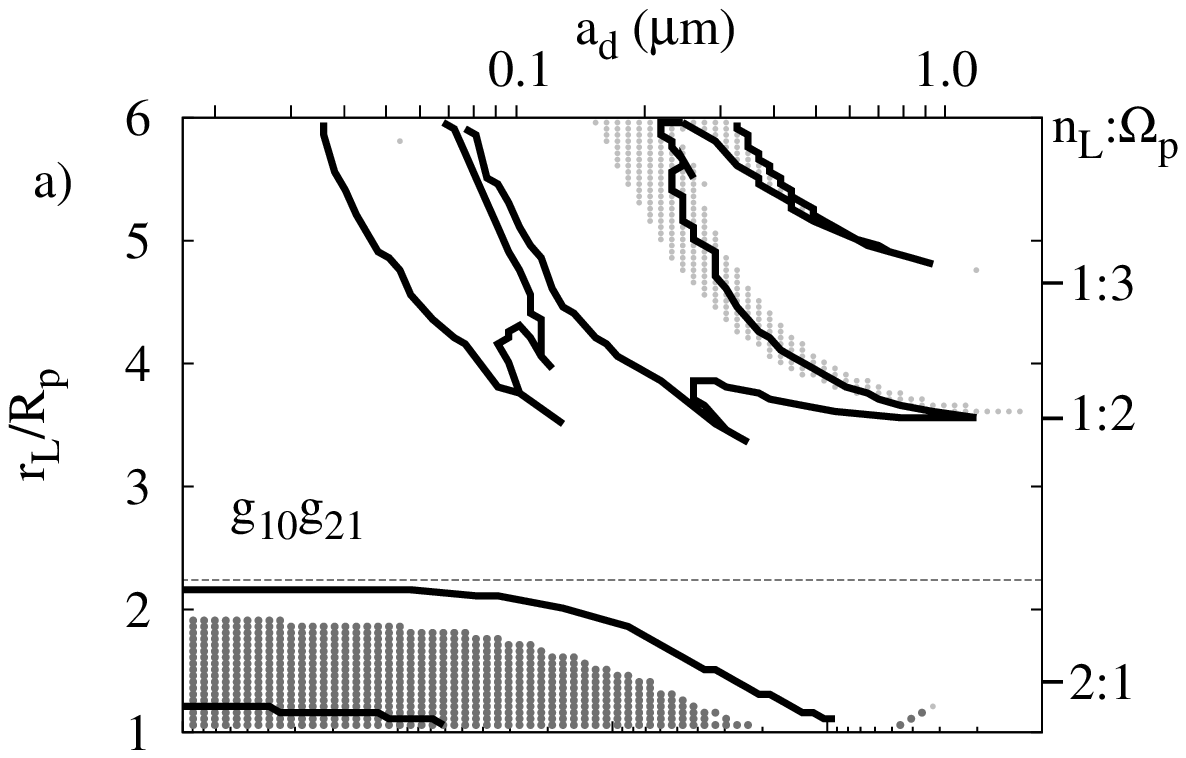}
\newline
\includegraphics [height = 2.1 in] {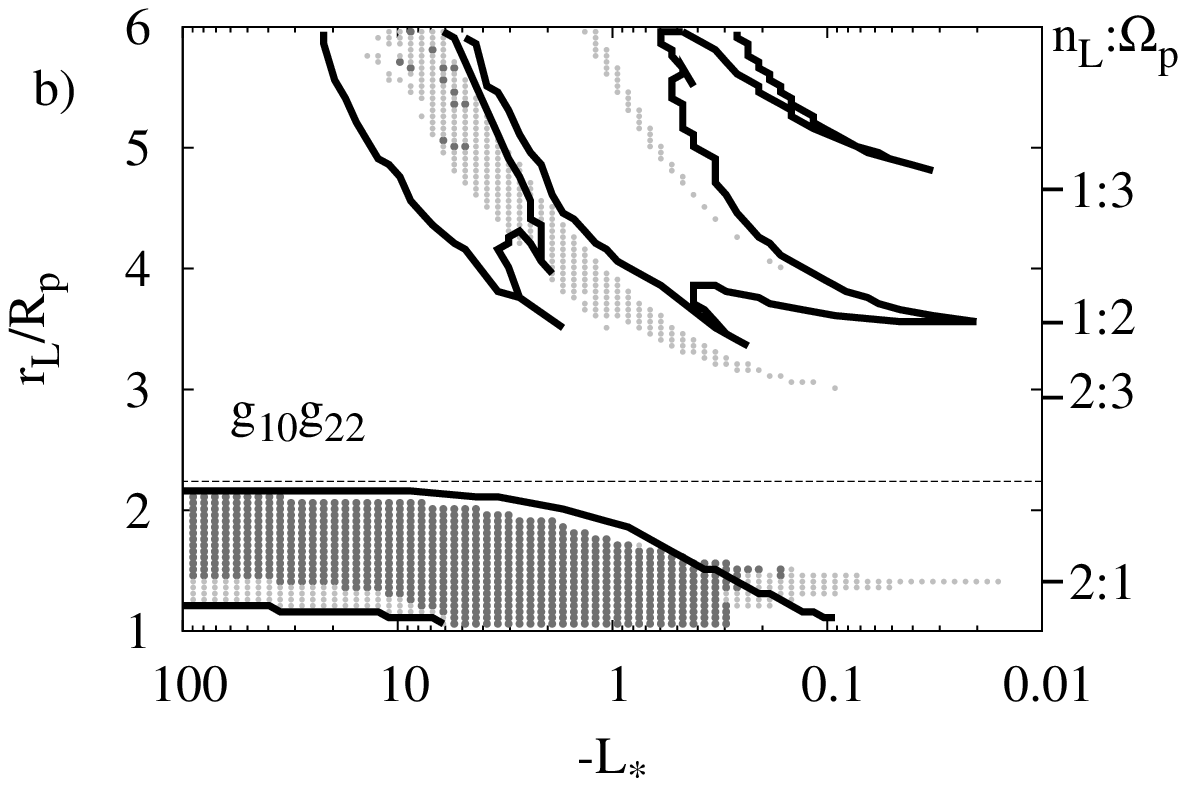}
\caption{\textbf{Quadrupole-order terms} isolated in stability maps
  for negatively-charged dust: To $g_{10}$ we add just $g_{21}$ in
  panel a) and just $g_{22}$ in panel b), with a 1-year simulation for
  each dust grain. The light grey marks stable grains whose radial
  excursions \textit{away} from $R_{syn}$ exceeded 4\% of the launch
  distance, while dark grey indicates collision just as in
  Fig.~\ref{fig:g10g11}a and~\ref{fig:g11res}a. The dark curves mark
  the envelope of instability when just the $g_{10}$ and $g_{11}$
  coefficients are included, from the numerical data of
  Fig.~\ref{fig:g10g11}a.}
\label{fig:g10g2neg}
\end{figure}

In this and the following sections, we focus on the escaping negative
grains outside synchronous orbit, because these escapes are the most
fundamental new effect added by a non-axisymmetric magnetic field. In
the stability maps of Fig~\ref{fig:g10g2neg}, we isolate the effects of
$g_{21}$ and $g_{22}$ to highlight their respective Lorentz
resonances, as compared to the stability boundaries of the tilted
dipole from Fig.~\ref{fig:g10g11}a (solid curves). For stable
negatively-charged grains in an aligned dipole field, radial motion is
always confined between the launch distance and synchronous
orbit. Thus, as in Fig.~\ref{fig:g10g11}a, the light grey data in
Fig.~\ref{fig:g10g2neg} trace where grain trajectories show motions
away from $R_{syn}$ that are significant on the scale of the launch
distance.

The $g_{21}$ and $g_{22}$ terms studied in Figs.~\ref{fig:g10g2neg}a
and~\ref{fig:g10g2neg}b clearly cause less overall instability than
$g_{11}$ in Fig.~\ref{fig:g10g11}a. Consider first, motion within
synchronous orbit. The $g_{21}$ term (Fig.~\ref{fig:g10g2neg}a) is
nearly as effective as $g_{11}$ in inducing vertical instability, and
in fact is better able to clear out the region just above the planet's
cloudtops. The effect of the $g_{22}$ term (Fig.~\ref{fig:g10g2neg}b)
is similar to, but typically weaker than, $g_{11}$ with one important
exception. Note the long, horizontal feature extending towards the 2:1
inner Lorentz resonance in the Kepler regime (small $\Lpar$). These
stable grains are strongly stirred by the 2:1 vertical Lorentz
resonance excited by $g_{22}$ (Table 1). A similar effect can be seen on the
right-hand side of Fig.~\ref{fig:g10g2neg}a; a small trail of points
near ($r_L/R_p \approx 1.05$, $\Lpar \approx -0.03$) hints that
$g_{21}$ causes a weak 2:1 inner Lorentz resonance. However, the first-order theory of \citet{ham94}
predicts that $g_{32}$ rather than $g_{21}$ should excite this resonance!

Outside $R_{syn}$, the situation is more
straightforward. Figure~\ref{fig:g10g2neg}a shows that the $g_{21}$
term strongly excites the 1:3 and 1:2 Lorentz resonances, as expected
from Table 1. Notice that the 1:2 resonance is significantly stronger
than the 1:3 resonance because, in the Kepler limit, the former has a
strength proportional to the small orbital inclination $i$ while the
latter's strength depends on the product of two small quantities $e$
and $i$, where $e$ is the orbital eccentricity
\citep{ham94}. Interestingly, in one way the $g_{21}$ term has a
stronger effect on radial motion than the $g_{11}$ term, extending the
1:2 and 1:3 resonances further into the Kepler limit. This can be
understood from Table 1 and \citet{ham94}, which show that the
$g_{21}$ term should naturally excite both of these resonances.

Figure~\ref{fig:g10g2neg}b shows two main features from the $g_{22}$
term outside $R_{syn}$, which tend towards 1:2 and 2:3 in the Kepler
limit. The $g_{22}$ term excites vertical motions \citep{ham94}, which are
not traced directly in Fig.~\ref{fig:g10g2neg}, but which clearly
couple to radial motions. This causes the outer 1:2 and 2:3 resonances
seen in Fig.~\ref{fig:g10g2neg}b, as well the strong inner 2:1
resonance that reaches far into the Kepler regime. A glance at Table 1
shows that $g_{22}$ excites a first-order 2:3 inclination resonance
and a second-order mixed and therefore weaker 2:4 resonance, accounting for the differing
responses of grains near these resonances visible in
Fig.~\ref{fig:g10g2neg}b.

We are left with a few paradoxes. First, how does the $g_{11}$
magnetic field term excite the 2:3 and 2:1 Lorentz resonance? And how
does the $g_{21}$ term excite the inner 2:1 resonance? To answer these
questions, we require a second-order expansion of the Gaussian
perturbation equations \citep{dan88}, for the electromagnetic force.
The first-order Fourier series expansion in the small parameter
$\Lpar$ was obtained by \citet{ham94} for each magnetic field
coefficient by treating the orbital elements as constants. To extend
this to second order, we take the Fourier series first-order solution
for each orbital element and insert it on the right-hand side of the
Gaussian perturbation equations.  Simplifying requires identities for
the product of two trigonometric functions and we end up with
second-order $\Lpar^2$ corrections to the time rates of change of the
orbital elements. Thus the power in each resonant frequency in Table
II of \citet{ham94} is augmented by a second-order
correction. Calculating the strength of these corrections is a
straight-forward but unenlightening exercise which we do not undertake
here, as the calculation is clearly invalid for $\Lpar > 1$ when the
third- and higher-order terms cannot be ignored. Indeed, the very
concept of orbital elements also breaks down for $\Lpar > 1$ when
electromagnetism is no longer a small perturbation to gravity.

Instead, we explore the form of the corrections and show how magnetic
field coefficients can excite resonant terms other than those shown in
our Table 1 and in Table II of \citet{ham94}. Consider first the
$g_{10}g_{11}$ simulation of Figs.~\ref{fig:g10g11}
and~\ref{fig:g11res}. To second order in $\Lpar$, this combination of
coefficients excites two relevant 2:3 resonances: $\dot{\Psi} =
3\omega_c -2\Omega_p -\dot{\varpi}_{peri}$ and $\dot{\Psi} = 3\omega_c
- 2\Omega_p + \dot\varpi_{peri} - 2\dot\Omega_{node}$, both with
amplitude proportional to $\Lpar^2 (R_p/r)^6 g_{11}^2 e i^2$. In
addition, a 2:1 resonance, $\dot{\Psi} = \omega_c - 2\Omega_p +
\dot\varpi_{peri}$ is also excited, with amplitude also proportional
to $\Lpar^2 (R_p/r)^6 g_{11}^2 e i^2$. These resonances show up in
Figs.~\ref{fig:g10g11} and~\ref{fig:g11res} near the planet where $(R_p/r)$ is relatively large, near instability boundaries where $e$
and/or $i$ are large, and for particles where $\Lpar$ itself is
relatively large. Comparison of the data in Fig.~\ref{fig:g10g11}b
 with the corresponding curves in Fig.~\ref{fig:g11res}b shows that these resonances are weaker than the
already-discussed first-order resonances, as expected.

In a similar manner, the second-order theory shows that $g_{21}$
also drives the 2:1 resonance with frequency $\dot{\Psi} = \omega_c
-2\Omega_p +\dot{\varpi}_{peri}$ and amplitude $\Lpar^2 g_{21}^2
(R_p/r)^8 e^3 $ (Fig.~\ref{fig:g10g2neg}a). The rest of
Figs.~\ref{fig:g10g2neg}a and~\ref{fig:g10g2neg}b appear to be well
explained by the linear theory of \citet{ham94}, which predicts both
pairs of instability outside $R_{syn}$.

The second-order corrections, however, should excite the resonant
frequency $\dot{\Psi} = 5\omega_c - 4\Omega_p + \dot\varpi_{peri} -
2\dot\Omega_{node}$ with amplitude proportional to $\Lpar^2 (R_p/r)^8
g_{22}^2 e i^2$; the tiny weak feature just below the 2:3 track and
near the center of Fig.~\ref{fig:g10g2neg}b may be due to this
resonance. Furthermore, the frequency $\dot{\Psi} = 3\omega_c -
2\Omega_p - \dot\varpi_{peri}$ with amplitude proportional to $L_*^2
(R_p/r)^8 g_{21}^2 e^3$, should excite particle motions in
Fig.~\ref{fig:g10g2neg}a, but no evidence for these motions is
seen. This may be due to the fact that horizontal resonances driven by $g_{21}$ are intrinsically weaker than the vertical resonances driven
by $g_{22}$ \citep{ham94}. In any case, given the strong drop in the
strength of second-order corrections with distance, their effects
outside $R_{syn}$ are minimal.
\subsection{Realistic Full Magnetic Field Models}

\begin{figure} [placement h]
\includegraphics [height = 2.1 in] {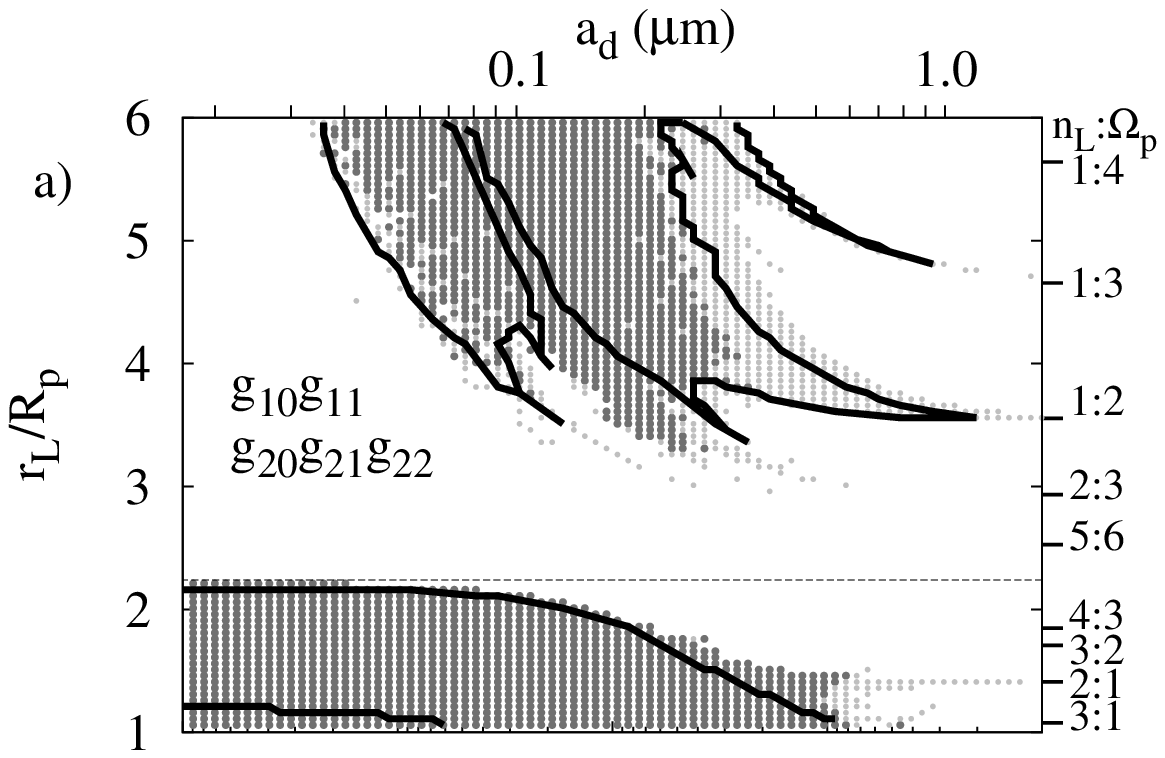}
\newline
\includegraphics [height = 2.1 in] {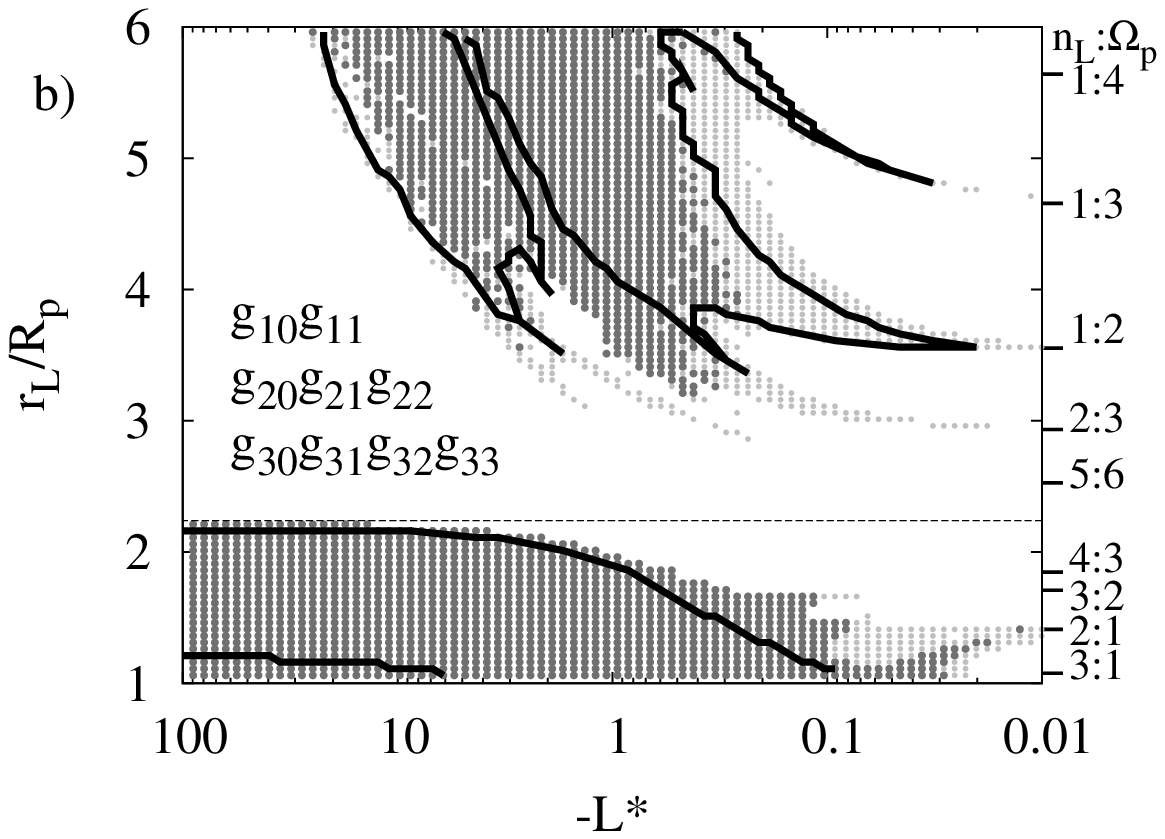}
\caption{Resonant \textbf{quadrupolar and octupolar effects} in stability maps
  for negative dust grains: To the tilted dipole model of
  Fig.~\ref{fig:g10g11}a, shown here as the solid curves, we add all
  quadrupole terms (panel a), and all terms out to octupole order
  (panel b). The darkest regions mark grains that collide with the
  planet or escape during a 1-year integration. The lighter grey
  indicates stable grains whose radial excursions \textit{away} from
  $R_{syn}$ exceeded 4\% of the launch distance, as in
  Figs.~\ref{fig:g10g11}a,~\ref{fig:g11res}a and~\ref{fig:g10g2neg}.}
\label{fig:all}
\end{figure}

Figure~\ref{fig:all}a combines the effects of all dipolar and
quadrupole terms for negative grains. Within synchronous orbit, all
grains in the Lorentz limit are now unstable, as are all grains within
the $g_{11}$ envelope. Furthermore, $g_{22}$ powers a 2:1 vertical
resonance with frequency $\dot{\Psi} = \omega_c - 2\Omega_p+
\dot\Omega_{node}$, seen as the dominant horizontal feature extending
from the instability region (inside $R_{syn}$ in Fig.~\ref{fig:all}a). Adding the octupole term $g_{32}$
strengthens this feature by exciting a 2:1 radial resonance $\dot{\Psi}
= \omega_c - 2\Omega_p + \dot\varpi_{peri}$ (Fig.~\ref{fig:all}b). These resonances have the
most dramatic effect on large dust grains near the planet, due to rapid decline in magnetic field strength with distance for higher-order terms. The $g_{33}$ octupole term adds a
spike of instability at the 3:2 resonance as well ($\dot{\Psi} =
2\omega_c - 3\Omega_p + \dot\Omega_{node}$).

Outside $R_{syn}$, a large region of escaping negative grains exceeds
the sum of the effects of $g_{11}$ (Fig.~\ref{fig:g10g11}a), $g_{21}$
(Fig.~\ref{fig:g10g2neg}a) and $g_{22}$ (Fig.~\ref{fig:g10g2neg}b),
although the main resonant tracks are easily identified.  In
particular, a huge swath of grains centered on $\Lpar =5$ at
$r_L/R_p = 5$ escapes here, but is bound for the simpler field
geometries of Figs.~\ref{fig:g10g11}a,~\ref{fig:g10g2neg}a, and~\ref{fig:g10g2neg}b.

Adding the octupole magnetic field coefficients (Fig.~\ref{fig:all}b)
presents only subtle differences from the quadrupole model of
Fig.~\ref{fig:all}a outside $R_{syn}$. In particular, the locations of
the resonant tracks appear to be unchanged. The three narrow fingers
in the center of Fig.~\ref{fig:all}b, however, are noticeably more
prominent than the corresponding structures in
Fig.~\ref{fig:all}a. The outer 2:3 resonance is driven by the $g_{22}$
term ($\dot{\Psi} = 3\omega_c - 2\Omega_p - \dot\Omega_{node}$), but
the 3:4 and 4:5 resonances cannot be excited by quadrupole terms in the
linear theory. The 3:4 resonance is driven by the $g_{33}$
coefficient, but also by a second-order term proportional to
$g_{11}g_{22}$. Both are active in the lower plot, while only the
latter affects the upper plot. Similarly, the 4:5 resonance is excited
by the non-linear $g_{22}^2$ term (both plots) and by the
$g_{11}g_{33}$ term (bottom plot only). As always, when multiple
resonances are active, chaos ensues and escape becomes more
likely. Note that these differences between Figs.~\ref{fig:all}a and~\ref{fig:all}b are confined within $\sim4 R_p$, due to the rapid
radial weakening of the high-order magnetic field terms.

In general, we see numerically and analytically that Lorentz resonances widen in strength as $\Lpar$ increases. This causes the
resonances to overlap and destabilize most of the grains near $\Lpar =
-1$ if grains are launched beyond the immediate vicinity of
synchronous orbit. As $|\Lpar|$ increases, higher-order dependencies
on the charge-to-mass ratio permit even more resonances to emerge and vie for control of dust grain dynamics.

\begin{figure} [placement h]
\includegraphics [height = 2.1 in] {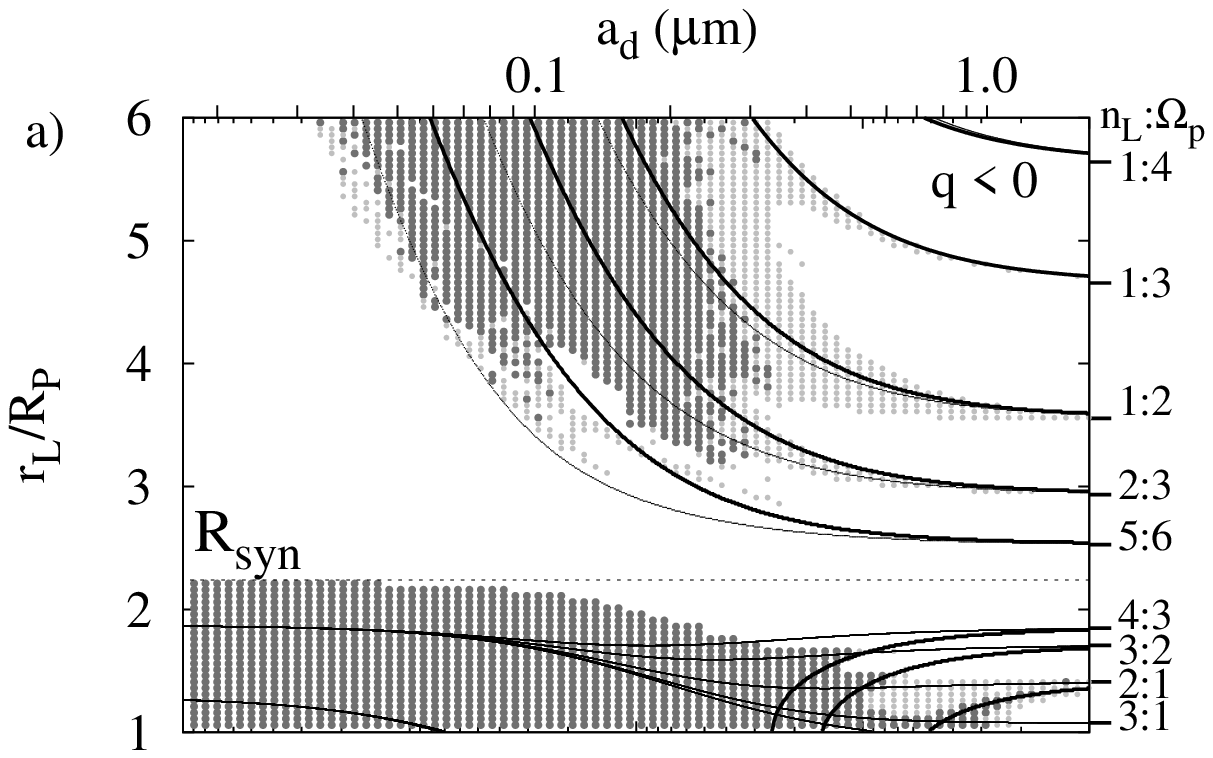}
\newline
\includegraphics [height = 2.1 in] {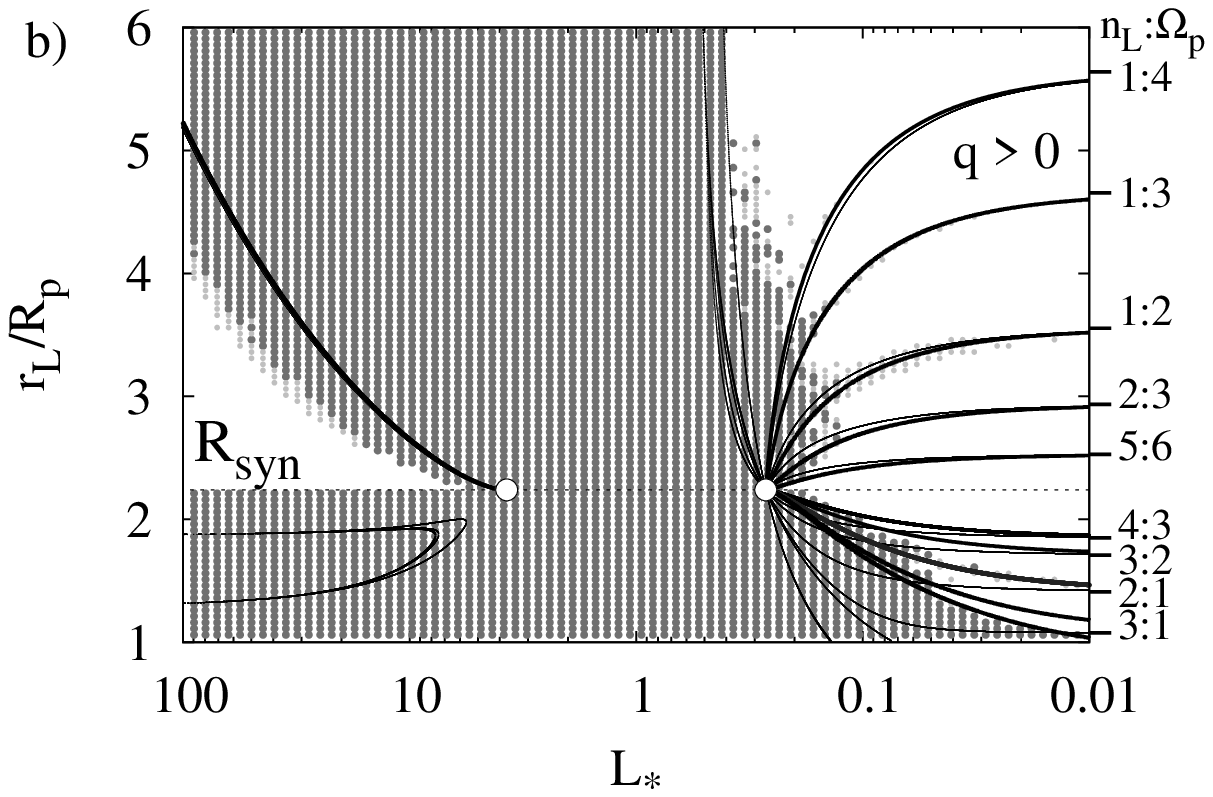}
\caption{Theoretical resonance curves over all charge-to-mass ratios superimposed on a stability map 
of Jupiter's full magnetic field for a) negative grains (from Fig.~\ref{fig:all}b) and b) positive grains. 
The dark grey marks dust grains that escaped or crashed into Jupiter during the 1-year integration. 
The faint grey points denote grains that experience radial motions away from synchronous that 
exceeded 4$\%$ of the launch distance for the negative grains, and towards $R_{syn}$ by at least 2$\%$ of $r_L$ for the positive grains.  
The thick bold curves mark radial Lorentz resonances, and the thin curves track the vertical Lorentz resonances. 
At synchronous orbit $\dot{\phi}_c = 0$, and the white circles mark the local stability threshold from \citet{jh12} where $\kappa_c \rightarrow 0$. }
\label{fig:fullfield}
\end{figure} 
In Fig.~\ref{fig:fullfield}, the vertical and radial Lorentz resonances for negative and positive grains are atop the stability map for Jupiter's full magnetic field modeled out to octupole order. For the negative grains of Fig.~\ref{fig:fullfield}a, as $\Lpar$ increases going from right to left, all the radial resonances diverge rapidly from $R_{syn}$. Most of the vertical resonances however, diverge from synchronous orbit more slowly as $|\Lpar|$ increases, and in the Lorentz regime these pile up on the vertical stability boundary inside $R_{syn}$ (Fig.~\ref{fig:1}a), where $\Omega_b$ $\rightarrow 0$ and hence, by the resonant condition, $\dot{\phi}_c \rightarrow 0$. The combined effects of many vertical resonances near this boundary destabilizes all grains in the Lorentz regime out to synchronous orbit in Fig.~\ref{fig:fullfield}a.

For the positive grains of Fig.~\ref{fig:fullfield}b, all radial resonances converge on the two locally unstable points along $R_{syn}$. In the Lorentz regime, the curve outside synchronous orbit satisfies $\kappa_c \rightarrow 0$. This is further to the left of the stability boundary for an aligned dipole (Fig.~\ref{fig:1}b), ensuring that resonances pile up and further destabilize grains with the additional magnetic field terms in the Lorentz regime. Physically, it means that smaller grains are more likely to be expelled for a particular positive electric potential than calculated using the aligned dipolar approximation. For the larger grains, in the Kepler regime, the vertical Lorentz resonances asymptote near the $\Lpar = 1/2$ boundary where the guiding center distance rapidly increases and the bounce frequency $\Omega_b \rightarrow 0$.   

For both positive and negative grains where $|\Lpar| <<1$, the outer 1:3 radial and vertical resonances coincide implying $\dot{\varpi}_{peri} = \dot{\Omega}_{node}$. This is indeed the case to first order in $\Lpar$ as was first deduced by \citet{ham93a}. The result can also be obtained from our Eqs.~\ref{omegac},~\ref{kappa} and~\ref{Omb}. 

In the Kepler regime, N:N+1 resonances pile up at synchronous orbit (like the 5:6 resonance marked in Fig.~\ref{fig:fullfield}a); for higher N, these are driven by $g_{NN}$ and $g_{NN-1}$ magnetic field terms beyond the octupole model that we have considered here. Thus our numerical model of Jupiter's magnetic field is incomplete and the inclusion of higher-order terms would lead to some additional escapes. Given the strong radial dependence of higher-order magnetic field components, we expect changes to be limited to regions close to the planet and at high $\Lpar$, just like the differences between Figs.~\ref{fig:all}a and~\ref{fig:all}b. Nevertheless, we eagerly await the improved magnetic field model that the Juno spacecraft will soon provide. In the meantime, we now relax our assumption of a constant grain-charge.

\section{Variable grain charge in an aligned dipolar field}
The electric charging of a circumplanetary dust grain is a function of the plasma
environment, the flux of solar radiation, and the physical properties
of the grain itself. Since the nature of the grains and their plasma
environment are poorly constrained, the motion of any particular grain
with a varying charge is highly model dependent. Our goal in this
section is not to pick the best model for a given situation, but
rather to elucidate the physics of orbital changes driven by charge
variations. Possibly the simplest non-trivial model which
nevertheless, must occur in circumplanetary applications is the
shutoff of the photoelectric current during planetary shadow passages
\citep{hb91}. This effect will be present even if all other
model-dependent charging effects are absent. The azimuthal asymmetries
that the shadow induces has a profound effect on dust grain motions,
as we shall soon see.

Returning to our aligned dipolar magnetic field model, and assuming
that the plasma distribution is perfectly axisymmetric, the effect of
the planetary shadow transit is to introduce a strong azimuthal
asymmetry in the charging environment for a dust grain. In the shadow,
the photoelectric effect of sunlight is absent, and interactions with
the plasma cause a net negative charge on the grains \citep{sb87}. In
the sunlight, by contrast, equilibrium typically favors a slight
positive electric potential.
\begin{figure} [placement h]
\includegraphics [height = 2.1 in] {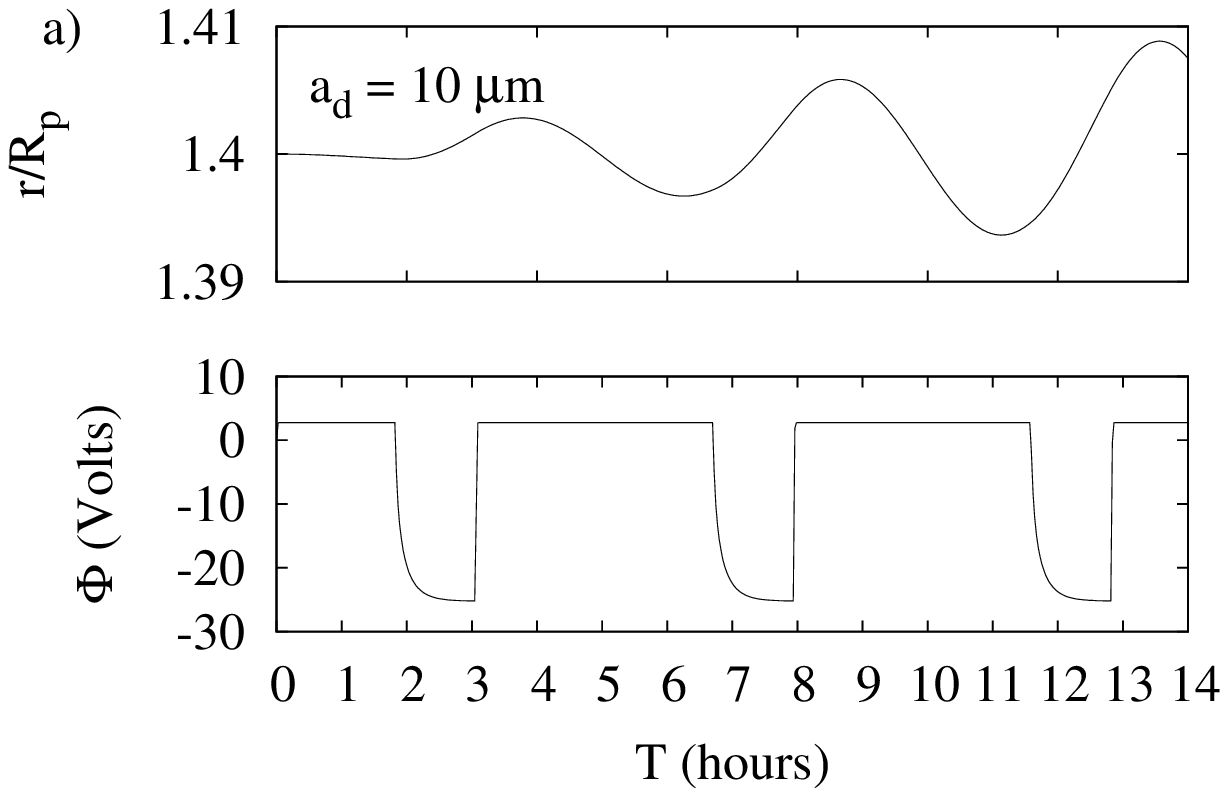}
\newline
\includegraphics [height = 2.1 in] {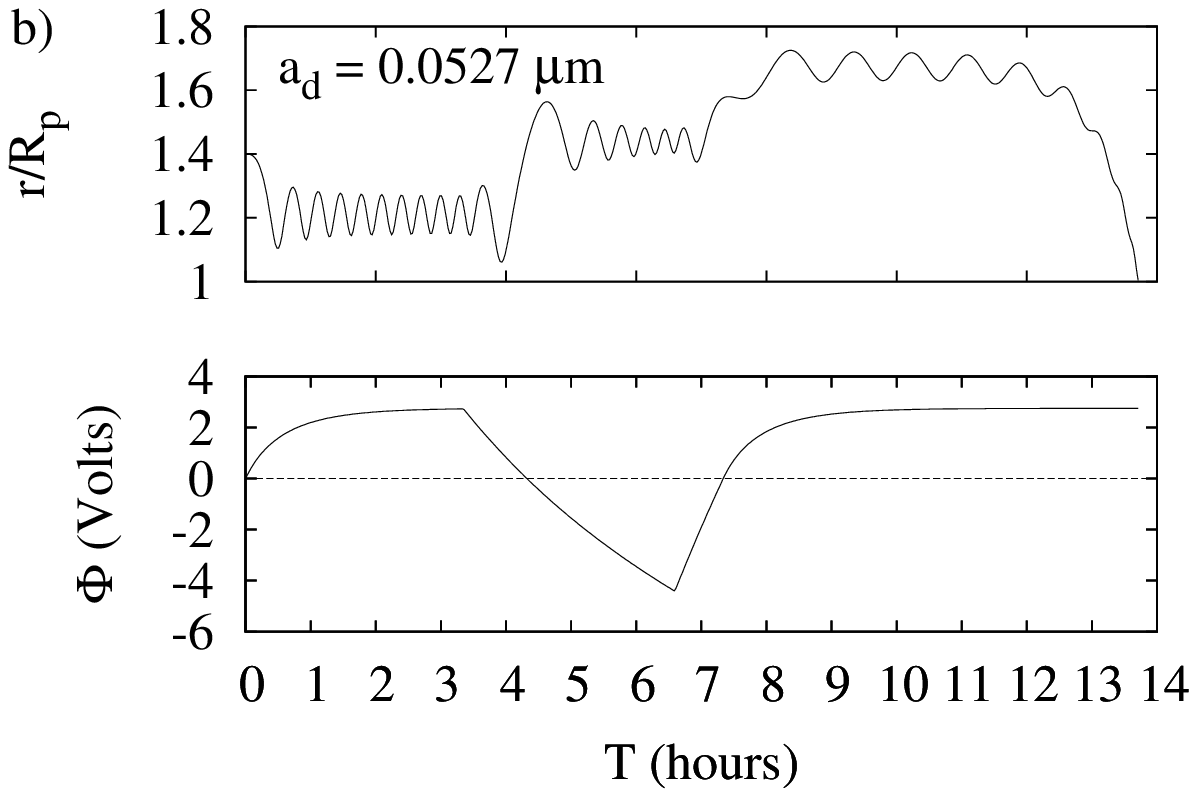}
\caption{The charge response of a dust grain in plasma and sunlight
  depends on grain size. Here we show a) 10 $\mu$m and b) 0.0527 $\mu$m grains,
  launched at 1.4 $R_p$ at local noon, with an aligned dipole magnetic
  field for Jupiter, in a uniform plasma with density $n_e = 1.4$
  cm$^{-3}$ and temperature $T_{e} = 10$ eV. The grains initially
  carry no charge. Each panel shows the grain's radial trajectory and
  instantaneous electric potential $\Phi$ in Volts. The large grain
  experiences three 1-hour long shadow passages during which the
  charge decreases, while the smaller grain has a single 3-hour
  eclipse, and much more sluggish changes to its electric potential.}
\label{fig:Lvar}
\end{figure}
Fig.~\ref{fig:Lvar} highlights the differing effect of charge
variation with grain size, given identical launch distances. Here, we
have chosen two example grains launched inside synchronous orbit at a
location that avoids the short term vertical instability of
Fig.~\ref{fig:1}.

Firstly, we note that the equilibrium charge on a sunlit dust grain is
2.75 Volts, whereas in the shadow, the equilibrium electric potential
is -27 Volts. The large grains reach their equilibrium potential far more
rapidly than the smaller grains. Indeed, the charge response is
typically inversely proportional to the size of the dust grain
\citep{hj10}. For the larger grain in Fig.~\ref{fig:Lvar}a, the
increasing amplitude of the radial oscillations is caused by the fact
that charge variation repeats each dust grain orbit, thereby
resonating with the epicyclic frequency for grains in the Kepler
regime. This is the destabilizing shadow resonance
(\citealt{hb91,hk08}) which we will find strongly affects our
stability map.

In Fig.~\ref{fig:Lvar}b, the smaller dust grain does not have enough
time to reach charge equilibrium during its three-hour shadow passage. This dust grain experiences stochastic kicks both radially inwards
and outwards from its launch distance; the grain eventually becomes
vertically unstable and crashes into the planet at high latitude after just
14 hours. Each kick in the guiding center distance $r_c$ occurs when
the electric potential on the dust grain is $\sim$1 Volt, when the
instantaneous value $\Lpar$ places the grain near the left-most radial
stability boundary of Fig.~\ref{fig:1}b. When the potential is higher
than 2 Volts or negative, the grain experiences stable radial
oscillations. The decreasing amplitude of these oscillations with time is due to the grain reaching higher charge-to-mass ratios
($|\Lpar|$), and hence experiencing tighter gyrations. After several
random steps in $r_c$ due to the grain's periodic encounters
with the radial instability, it moves into the vertical instability zone (Fig.~\ref{fig:1}) and is lost to Jupiter.

\begin{figure} [placement h]
\includegraphics [height = 2.1 in] {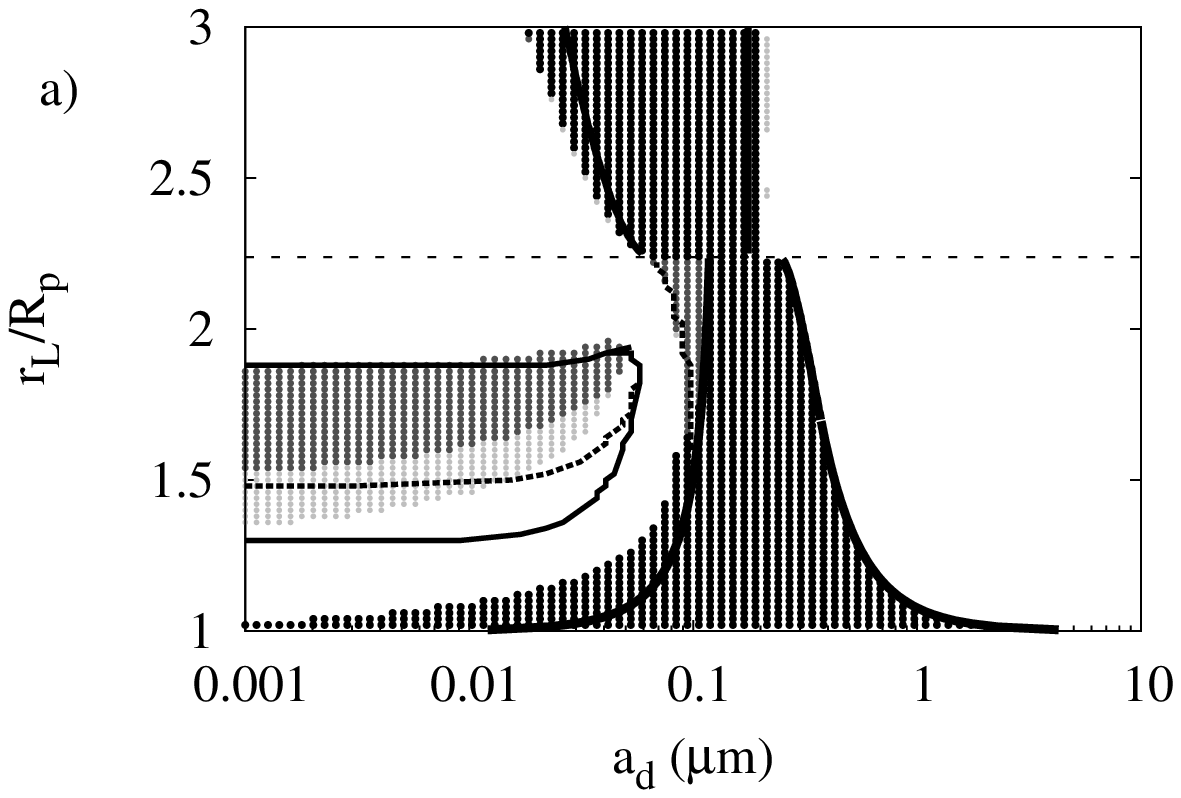}
\newline
\includegraphics [height = 2.1 in] {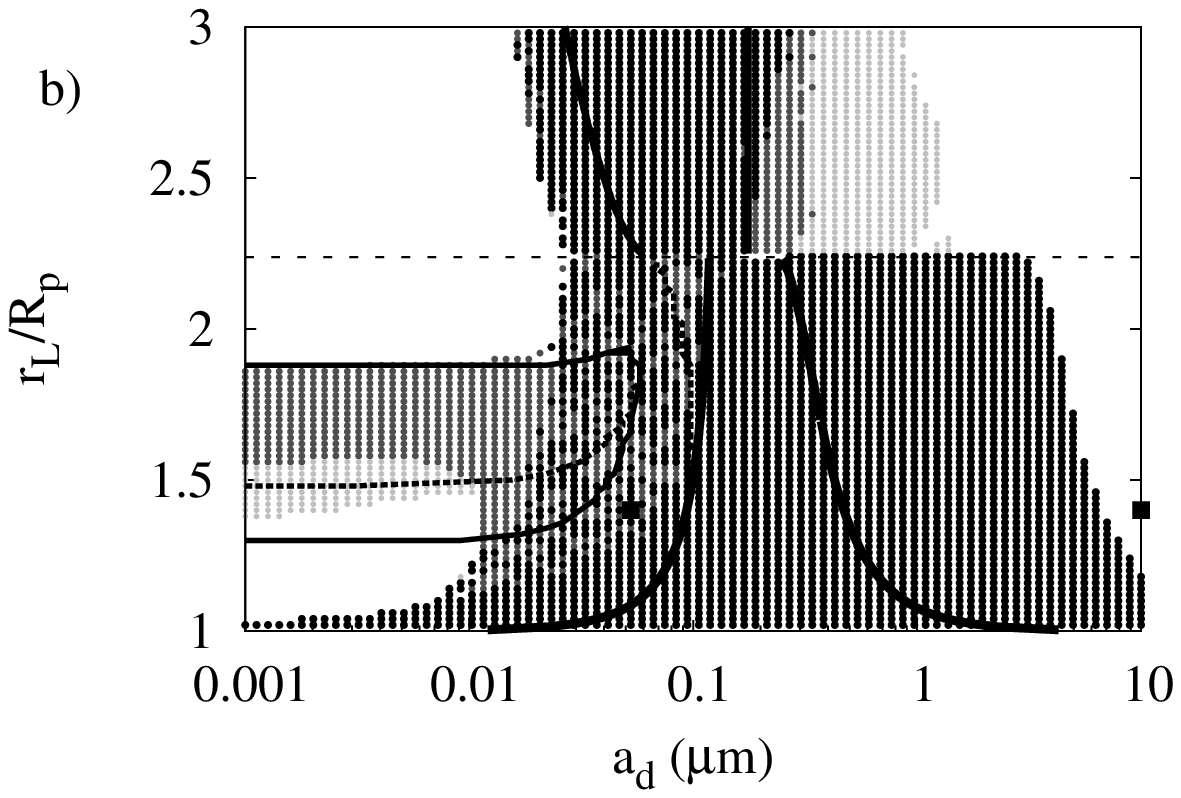}
\caption{Stability of grains with variable charge response to plasma
  for two grain charging models: a) constant photoelectron emission,
  and b) a more realistic photoelectric response which is interrupted
  during shadow transit. We adopt an aligned dipole for Jupiter's
  magnetic field, and neglect its 3.12$^{\circ}$ obliquity. Grains,
  all launched at local noon, begin with zero electric potential and
  are integrated over 1 year, a timescale long enough to cover the
  orbital precession in the Kepler regime due to $J_2$, which is
  included in our model of the gravity field. Consistent with the greyscale in
  Figs.~\ref{fig:Lmapvvar} and ~\ref{fig:g20negpos}, the darkest
  region denotes radial instability- either collision with the planet
  or escape within $\lambda_m = 5^{\circ}$ of the equator plane- the
  moderate grey represents high-latitude collisions with Jupiter, and
  the lightest grey marks surviving grains with high latitude
  oscillations. The two filled squares correspond to the trajectories
  illustrated in Fig.~\ref{fig:Lvar}. Note that the horizontal axis,
  now marking increasing grain radius as $\Lpar$ is no longer constant, spans twice the range as in
  previous figures. In both a) and b), the numerical boundaries are
  included for an assumed electric potential of +2.75 Volts. }
\label{fig:JUPchargevary1}
\end{figure}

In Fig.~\ref{fig:JUPchargevary1} we present stability maps for a large
range of grains sizes from 0.001$\mu$m to 10$\mu$m, over a broad range
of launch distances, to test the effect of charge variations on
orbital stability. In these models, dust grains are free to vary their
charge as the environment allows, both with the effects of the shadow present
(Fig.~\ref{fig:JUPchargevary1}b), and explicitly ignored (Fig.~\ref{fig:JUPchargevary1}a).

Without the planetary shadow, grain charges quickly converge to
equilibrium values, and the stability map in
Fig.~\ref{fig:JUPchargevary1}a looks very similar to one for a
constant (positive) charge (Fig.~\ref{fig:1}b). In
Fig.~\ref{fig:JUPchargevary1}a, the superimposed bold-faced curves,
corresponding to a +2.75 Volt constant potential, match the data very
closely on far right. Since the electric potentials of the large grains rapidly converge to equilibrium, the
Kepler-regime side of the radial instability also closely conforms to the
analytical boundaries of \citet{jh12}. The smaller grains, however,
take significant amounts of time to reach charge equilibrium. The
grains just to the left of the left-most solid curves in
Fig.~\ref{fig:JUPchargevary1}a either escape (outside $R_{syn}$) or
fall into the planet (inside $R_{syn}$), before they have enough time
to reach their equilibrium charge. While these tiny grains experience modest electric charges, a
different set of stability curves to the left of those in Fig.~\ref{fig:JUPchargevary1}a applies. Within
1.2 $R_p$, for example, 0.01 $\mu$m sized grains collide with Jupiter within a few
hours while the characteristic charging time is a day. Similarly,
outside synchronous orbit, 0.01 $\mu$m-sized grains just outside the
radially unstable zone for constant +2.75 Volt grains can still escape the
planet. These grains, initially neutral, charge up slowly in the
sunlight. Hence, even if their equilibrium charge would permit stable
motion, the time spent in the radially unstable regime causes them to
collide with Jupiter or escape before reaching charge equilibrium.

With the planetary shadow turned on, the shadow resonance acts to
increase eccentricities, destabilizing grains over a far broader range of sizes than than those that remain at their equilibrium
potential. For the largest of these grains ($a_d\gtrsim 1 \mu m$), the shadow transit destabilizes
grains over timescales commensurate with the precession due to the
gravitational $J_2$ term \citep{hb91}. Thus we include the $J_2$ term
in our model for Fig.~\ref{fig:JUPchargevary1}, and adjust launch
speeds to ensure launch from a circular orbit. The smallest grains in
Fig.~\ref{fig:JUPchargevary1}b respond to changes in the charging
environment over a longer timescale. Thus, grains that survive the
initial charging process (the stable grains of
Fig.~\ref{fig:JUPchargevary1}a) reach an electric potential that
deviates little from its mean over the orbital period, and hence the
shadow has little effect on grains smaller than 0.01$\mu$m in
size. This region of Fig.~\ref{fig:JUPchargevary1}b essentially
matches Fig.~\ref{fig:1}b in the Lorentz limit, with either vertical
instability or stable high-latitude oscillations between 1.29 $R_p$
and 1.70 $R_p$.

Grains between 0.01$\mu$m and 0.1$\mu$m launched outside 1.2 $R_p$,
but within $R_{syn}$, experience charge variations that cause them to
spend some fraction of each orbit in a radially unstable
regime. Eventually they strike the planet, although the timing for
this is unpredictable. As we saw in Fig.~\ref{fig:Lvar}b, such grains
experience random walks in radial location but do not cross
synchronous orbit. Roughly half the grains in this region of
Fig.~\ref{fig:JUPchargevary1}b collided with the planet at high latitudes.

Grains larger than 0.5$\mu$m launched outside $R_{syn}$ in
Fig.~\ref{fig:JUPchargevary1}b experience excited radial motions and
vertical motion close to the radial stability boundary of
Fig.~\ref{fig:JUPchargevary1}a ($\Lpar = \frac12$,
\citealt{ham93a,jh12}), which extends the Thebe ring away from Jupiter
\citep{hk08}. Inside $R_{syn}$, Fig.~\ref{fig:JUPchargevary1}b shows
that the shadow resonance destabilizes grains more than 10 times bigger than the largest grains destabilized with the shadow
switched off. The boundary between stable and unstable here is
determined not by the time of the integration but by the precession
timescale due to the higher-order $J_2$ gravity field component at Jupiter
($\approx$ 0.25 years in the main ring, and longer further out).

We emphasize that resonant charge variation on dust grains due to the
photoelectric current clearly has an important effect on grain
dynamics. Epicyclic motion also provokes resonant charge variations
due to both radial gradients in the plasma properties and the varying
dust-plasma speed with gyrophase. Of the effects studied so far:
launch speeds, realistic magnetic fields and time-variable electric
charges, the latter appears to be the most important. Non-zero launch
impulses, by contrast, are a minor effect on grain-orbit
stability. The relative importance of the different effects, however,
will vary dramatically with plasma properties.

For example, in a dense plasma like the Io plasma torus, the
equilibrium charge is always negative (as in Figs.~\ref{fig:g10g11}a and~\ref{fig:g11res}a) even in full sunlight. Spatial or
temporal gradients in plasma properties can also have a profound
effect. Since all of the parameters for charging are very uncertain,
we leave a thorough study of these effects for another paper. The constant charge maps here and in \citet{jh12}
are still relevant though, serving as a measure of the minimum
instability in a given system. It is also a good approximation for planets
with significant obliquities, where dusty rings spend much of their host's long orbital
period in direct and uninterrupted sunlight.

Accordingly we turn to the other major magnetospheres of the Solar System
and construct stability maps for dust grains with constant charge in
complex multipolar magnetic fields.
\section{Other planets}
After our detailed investigation of Jupiter, we are now in a position
to map and interpret stability results for each of the magnetized planets in the
Solar System. We begin with the planet with the simplest magnetic
field, Saturn.
\subsection{Saturn}
Saturn's full magnetic field can be described by an aligned dipole
with a slight vertical offset \citep{cao11}.
\begin{figure} [placement h]
\includegraphics [height = 2.1 in] {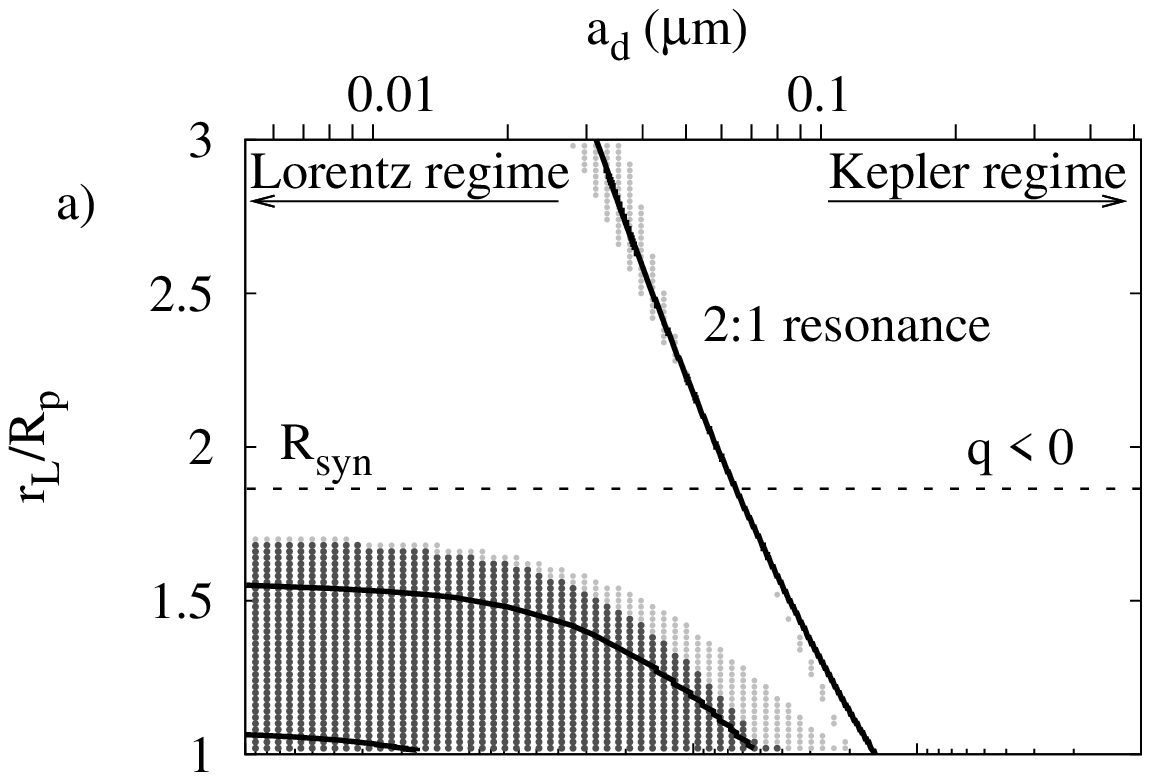}
\newline
\includegraphics [height = 2.1 in] {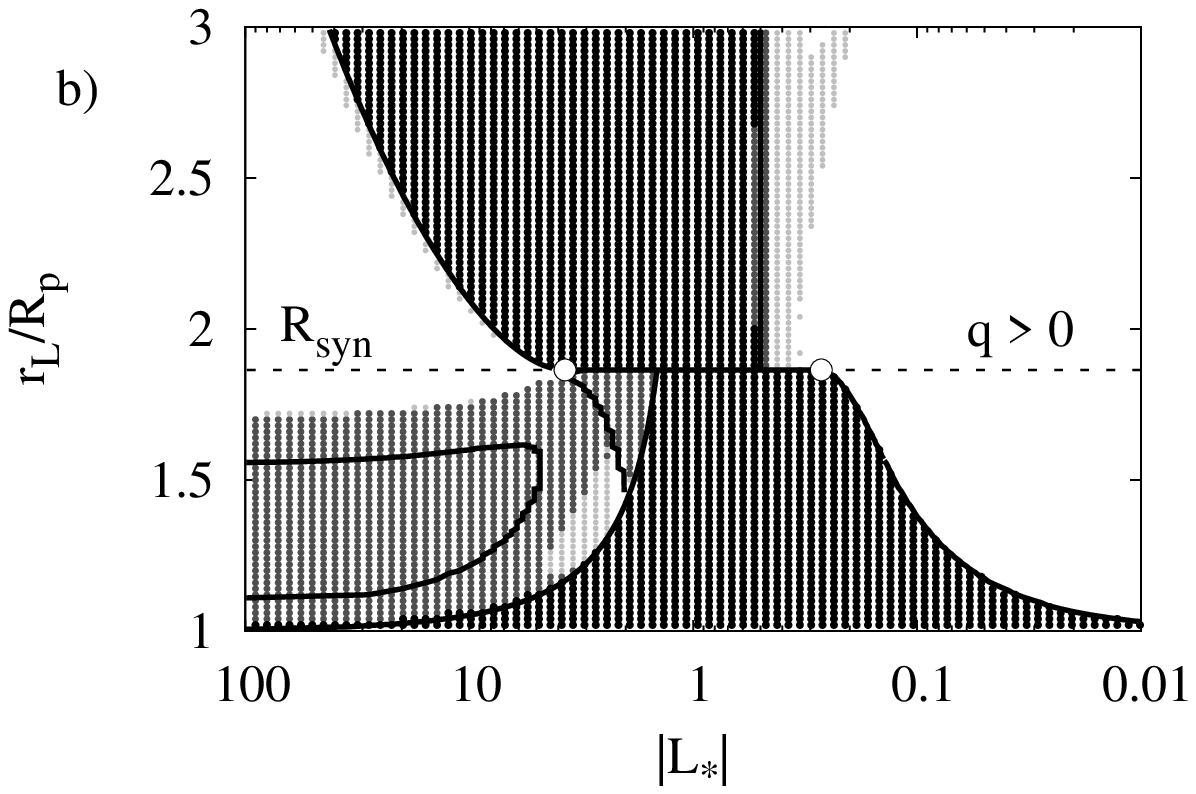}
\caption{Stability of a) negative and b) positive Kepler-launched
  grains, followed for 0.1 years in Saturn's full magnetic field. The
  greyscale matches that of Figs.~\ref{fig:Lmapvvar} and~\ref{fig:g20negpos}. The darkest region denotes grains that are
  radially unstable to escape or strike the planet (positive charges
  only), the moderate grey scale marks trajectories that were
  vertically unstable to climb out of the ring plane and strike the
  planet at high latitude, the light grey scale marks grains that are
  vertically unstable in the equatorial plane but remain globally
  bound with mirror points further than $|\lambda_m| = 5^{\circ}$ from
  the equator, and the white area marks locally stable orbits.
  Superimposed on the data are the numerically-determined stability
  boundaries for an aligned and centered dipolar magnetic field model
  for Saturn \citep{jh12}, as well as the curve marking the analytical
  2:1 resonance between epicyclic and vertical motion, which closely
  tracks band of grains that reach high latitudes but remain bound. The white circles mark the local stability threshold at $R_{syn}$, 
  where $\Lpar=2 \pm\sqrt{3}$.}
\label{fig:SATg10g20g30}
\end{figure}
We model the full field with $g_{10} = 0.2154$, $g_{20} = 0.0164$ and
$g_{30} = 0.0274$ Gauss \citep{cdc84}. Figure~\ref{fig:SATg10g20g30}a
shows the stability map for negatively-charged grains, with the
numerically-determined stability boundaries for an aligned and
centered dipole included for comparison, as in
Fig.~\ref{fig:g20negpos}.

As at Jupiter, (see Fig.~\ref{fig:g20negpos}) Saturn's dipole offset
increases the instability of grains to vertical perturbations. This
eliminates the stable zone close to the planet that we see for the
aligned dipole case, and moves the outer vertical stability boundary
significantly further from the planet. The effect is stronger at
Saturn than at Jupiter due to its relative large $g_{20}$ term and to the larger
$R_p/R_{syn}$ at Saturn, making the planet a bigger target. By
contrast, at Jupiter (see Fig.~\ref{fig:g20negpos}), a large locally-stable
region in the Lorentz limit close to the surface and at high $|\Lpar|$ survives the
inclusion of $g_{20}$.

For positively-charged grains, in Fig.~\ref{fig:SATg10g20g30}b, the
offset dipolar field causes the vertical instability to join the
radial instability, as in Fig.~\ref{fig:g20negpos}b. However, a tiny
island of globally stable grains survives near ($\Lpar = 3$, $r_L/R_p
= 1.3$).  The radial stability boundaries for an aligned dipole field
for Saturn \citep{jh12}, match Saturn's full magnetic field remarkably
well.

At Saturn we also see a slightly wider range of charge-to-mass ratios
excited by the 2:1 resonance between epicyclic and vertical motions,
when compared to Jupiter (Fig.~\ref{fig:g20negpos}a). This is due to
the range of launch distances extending further out in units of
$R_{syn}$ in Fig.~\ref{fig:SATg10g20g30}a. The accuracy of the theoretical curve matching this resonance vindicates the use of an aligned dipole approximation for Saturn's magnetic field to calculate radial and vertical orbital frequencies.

The transition from grains that are lost to the vertical instability
to those that remain bound in the B ring is at $1.70 R_p$ or $\approx
102,000$ km in the Lorentz limit (Fig.~\ref{fig:SATg10g20g30}). This
is close to a large increase in optical depth in the B ring that begins
around 1.72$R_p$, and losses to erosion may play a role in ring
evolution across this boundary. \citet{nc87} argued for a link between
the inner edge of B ring and the vertical stability boundary. Their
model for vertical motion predicted all highly-charged grains to be
unstable within 1.54$R_p$ at Saturn, close to the sharp inner edge of
the B ring. Their
model, however, did not provide a detailed mechanism and also invoked poorly understood electrostatic effects
(\citealt{nh83b,nc87}). Voyager 2 data revealed another transition in
optical depth between 1.63 $R_p$ and 1.65 $R_p$ (98,000 - 99,000 km),
a few thousand kilometers inside the vertical stability boundary at
Saturn with its full magnetic field configuration. While the proximity
of this transition to the vertical stability boundary is
intriguing, a detailed model to explain this congruency remains
elusive.

\subsection{Earth}
The Earth's magnetic field is dominated by a dipole tilted by a
moderate 11.4$^{\circ}$ from the axis of rotation. For our full-field
models, we use magnetic field coefficients out to octupole order from
\citet{rs72}. For the Earth, $g_{10} < 0$ and the magnetic field is
inverted compared to all of the giant planets. Thus for the Earth,
$\Lpar > 0$ for negatively-charged grains, and it is these negative
grains that suffer the radial instability \citep{jh12}.

\begin{figure} [placement h]
\includegraphics [height = 2.1 in] {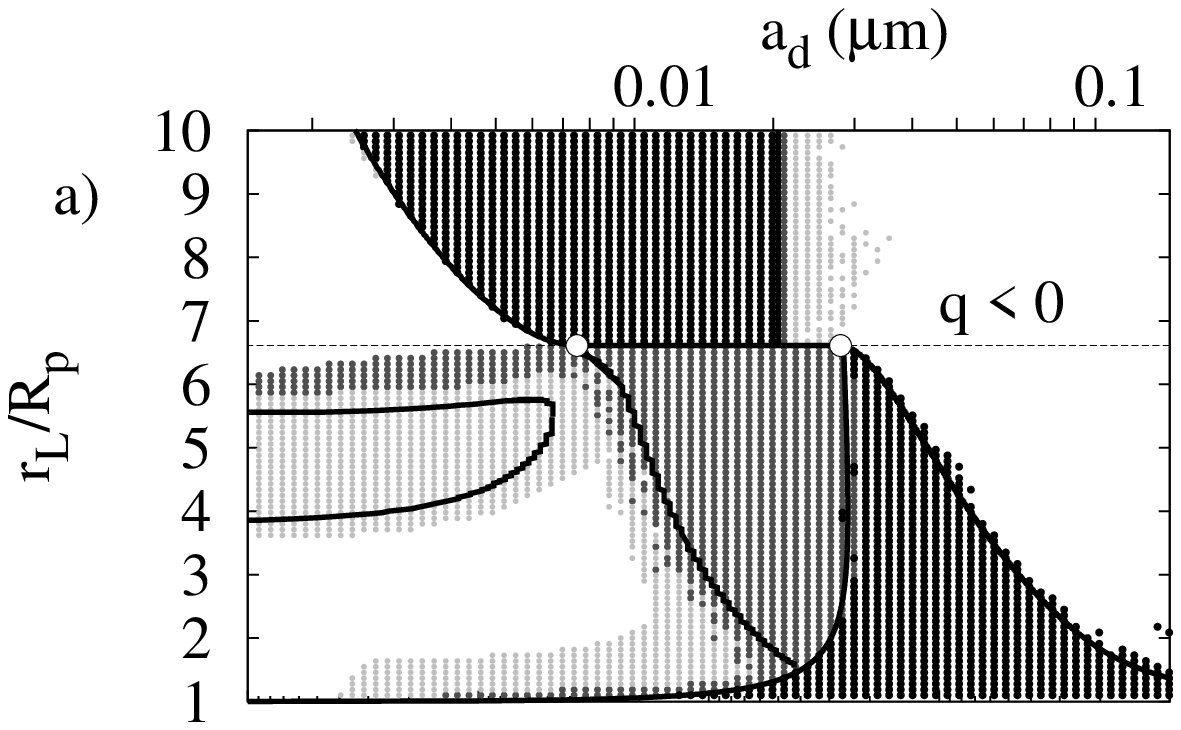}
\newline
\includegraphics [height = 2.1 in] {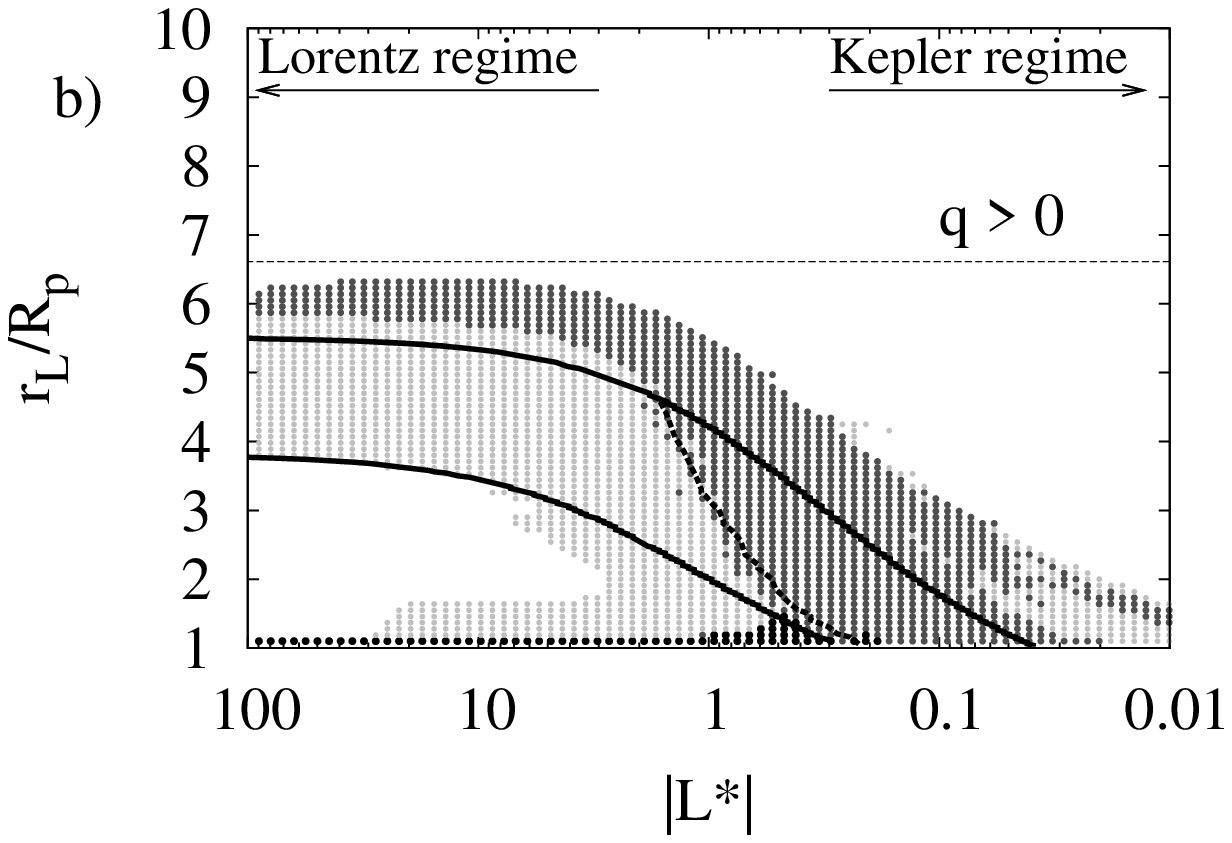}
\caption{Stability of Kepler launched negative (a) and positive (b)
  grains in Earth's magnetic field complete to octupole order,
  integrated over 1 year, with launch at $\phi_0 = 0$. The solid
  curves mark the numerical stability boundaries for the Earth with
  its anti-aligned $g_{10}$ component alone (Fig. 12 from
  \citealt{jh12}). Grains in the darkest region, (negative charges
  only) crashed into the planet or escaped at a latitude less than
  twice Earth's 11.4$^{\circ}$ tilt angle. The moderate grey region
  marks grains that struck the planet at latitudes higher than twice
  the tilt angle, the light grey marks grains with stable vertical
  oscillations with mirror points exceeding twice the tilt angle in
  latitude, and white regions are the remaining stable orbits. 
  The white circles mark the points $\Lpar = 2\pm\sqrt{3}$, at $r_L = R_{syn}$, where grains are on the threshold of radial instability.}
\label{fig:EAR}
\end{figure}

Figure~\ref{fig:EAR} compares the stability of grains in Earth's full
magnetic field to an aligned dipolar model. With an aligned dipolar
field, the vertical instability at Earth in the Lorentz limit is local, leading to a
region of high-latitude globally-stable oscillations. For the positive grains ($\Lpar < 0$, Fig.~\ref{fig:EAR}b)
this locally vertically unstable region curves towards the planet as
$|\Lpar|$ decreases, and only a small range of grains launched near
the surface between $\Lpar \approx -0.1$ and $-1.0$ collides with the
planet. This changes very little with the inclusion of Earth's
higher-order magnetic field terms, as Fig.~\ref{fig:EAR}
indicates. More dramatically, the higher-order terms (primarily
$g_{11}$) expand the vertical instability in the Lorentz limit to
further distances, almost to $R_{syn}$. In this region, grains do in
fact collide with the planet. The feature is very similar to what we
saw at Jupiter in Figs.~\ref{fig:g11fineg} and~\ref{fig:g11fipos}, except that the global instability region is far narrower in the case of the Earth. Just as
at Jupiter, tilting the magnetic field does not significantly move the inner
vertical stability boundary of Fig.~\ref{fig:EAR}b. For the negative grains,
Fig.~\ref{fig:EAR}a shows that the vertical instability is also displaced
towards $R_{syn}$ and curves slightly upwards to merge with the region
of radial instability. The expansion of the vertical instability for both positive and negative charges nearly to synchronous orbit has a unique 
benefit in assisting the removal of dusty space debris from this crowded region of Earth orbit (\citealt{hhm88,jh97,vl08}).

The radial instability in Fig.~\ref{fig:EAR}a, however, looks very different at Earth than at
Jupiter (compare Figs.~\ref{fig:g11fipos} and~\ref{fig:EAR}a). Nevertheless, outside synchronous orbit, the higher-order magnetic field
terms have little effect and the radial stability boundaries of
\citet{jh12} match the data remarkably well. Inside synchronous orbit,
the more complicated magnetic field slightly extends the area of HRLOs that abut the disjoint regions of
equatorially-confined radial instability. Thus Earth's full magnetic
field barely alters the radial instabilities expected for an aligned
dipole: The few unstable grains that exceed the radial stability
boundary on the right side of Fig.~\ref{fig:EAR}a may be associated
with Lorentz resonances. In particular, the small cluster of points
near ($\Lpar = 0.01$, $r_L/R_p = 2.0$) closely corresponds to the
inner 6:1 Lorentz resonance. All else being equal, the Lorentz
resonances are more important close to the planet where the magnetic
field irregularities are strongest.

As an exercise, we compared the results displayed in
Fig.~\ref{fig:EAR} with a simpler tilted dipole model, including just
the $g_{10}$ and $g_{11}$ terms (figure not shown). The main difference that
arises is that the positive and negative grains that are excited to
high latitudes, near ($|\Lpar| = 10$, $r_L/R_p = 1.3$) are not excited
in the tilted dipole model. A more subtle difference is the extra set of collisions to the right of the radial instability boundary on the right-hand side of Fig.~\ref{fig:EAR}a marking grains that were lost because of Lorentz resonances. These grains survive in the simple tilted dipole model. Deviations at greater distances are not
expected due to the steep radial dependence of the quadrupole and
octupole terms, and indeed, they are not seen. All in all, a tilted
magnetic dipole is a robust model for the motion of charged dust
grains at the Earth.
\subsection{Uranus} 
\begin{figure} [placement h]
\includegraphics [height = 2.1 in] {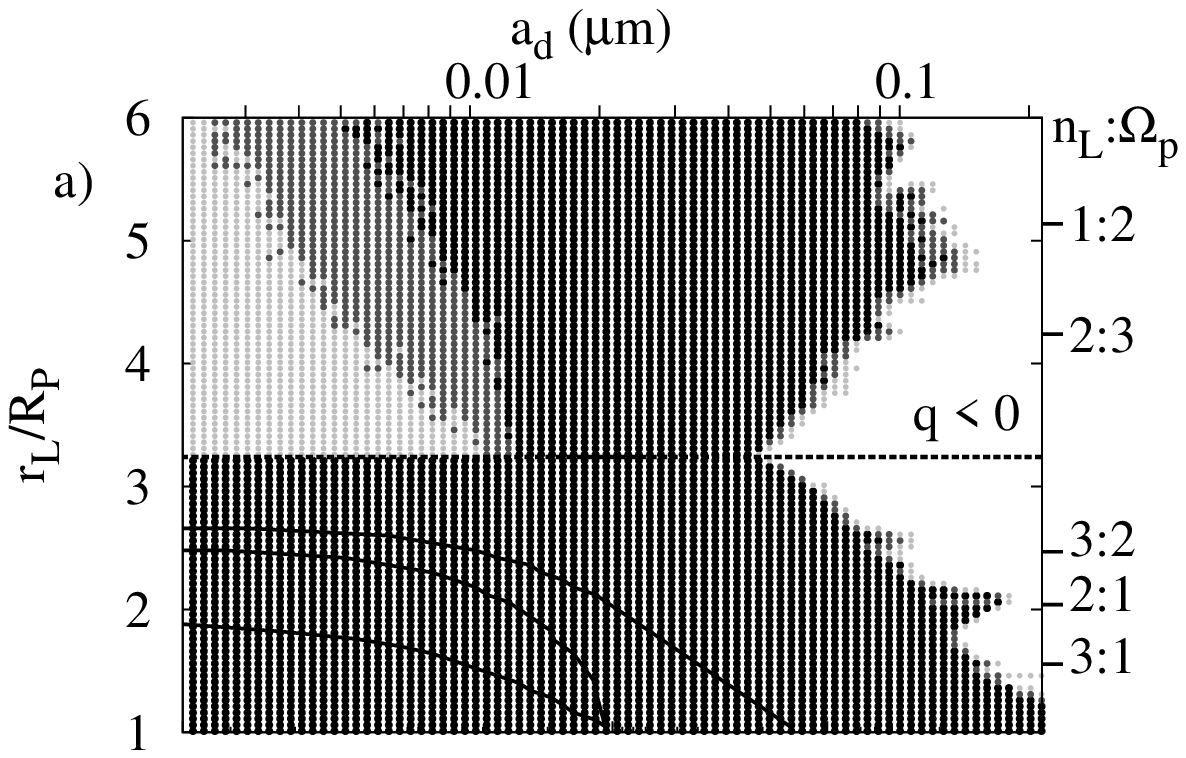}
\newline
\includegraphics [height = 2.1 in] {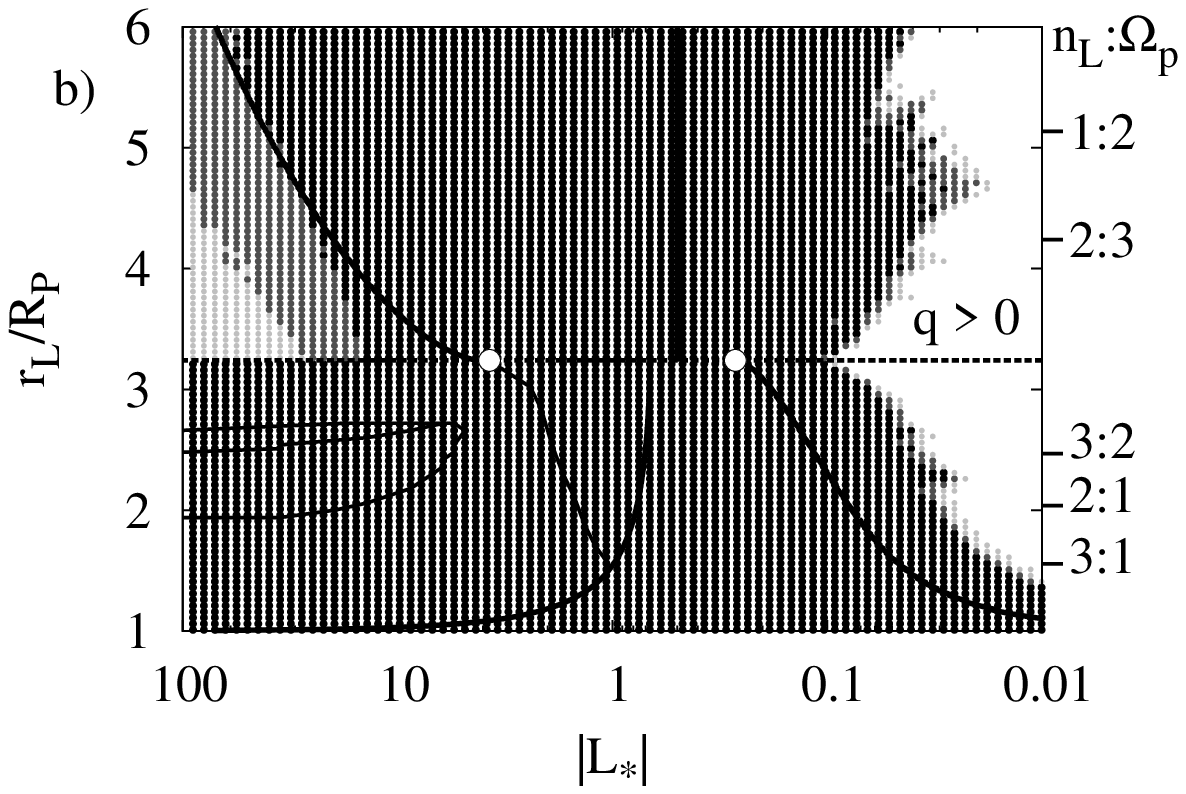}
\caption{Stability of Kepler-launched grains in Uranus' full field
  over one year. There are three shades of grey, plus white to
  highlight the effect of azimuthal launch position on grain orbit
  stability. The darkest grey marks grains that were unstable for all
  of 12 equally-spaced launch longitudes. The intermediate grey
  denotes unstable trajectories for 6 to 11 launch positions, the
  lighter grey 1 to 5 launches, and the white regions were stable for
  all launch positions. The embedded curves mark the Uranian equivalent of those in Fig.~\ref{fig:1}, 
  the stability boundaries for an aligned dipole magnetic field model. 
  The white circles mark the local radial stability at $R_{syn}$, where $\Lpar = 2\pm\sqrt{3}$.}
\label{fig:URA12}
\end{figure}

Uranus' complex magnetic field destabilizes grains for a much wider
range of charge-to-mass ratios than Jupiter, Saturn or
Earth. Figure~\ref{fig:URA12} shows the stability of grains launched
at Uranus, with magnetic field coefficients out to octupole order
taken from \citet{nes91}. This figure highlights the significant
dependence of launch azimuth on grain lifetimes. We determined grain orbit stability for 12 equally-spaced azimuthal launch positions, and followed
trajectories for 1 year.

Beyond $R_{syn}$, the full Uranian magnetic field causes a
large class of grains to escape rapidly, for both negatively- and
positively-charged dust. In general the stability maps for negative
and positive grains are very similar, especially inside synchronous
orbit.  Within $R_{syn}$, all trajectories in the Lorentz limit appear
unstable for both negative and positive charges, which significantly
constrains the low-energy plasma environment in the uranian ring
system. Furthermore, both Figs.~\ref{fig:URA12}a and ~\ref{fig:URA12}b
show far more dependence on launch phase at high charge-to-mass ratio,
on the left side of the stability maps, than on the right, consistent
with our results for Jupiter (Fig.~\ref{fig:g11fipos}).
  
Uranus' magnetic tilt as well as its quadrupole and octupole magnetic
field coefficients are much more important when compared to its
$g_{10}$ term than at Jupiter or Saturn, causing strong Lorentz resonances,
and hence a dramatic expansion of escaping negative grains over that
seen at Jupiter in Fig.~\ref{fig:all}b. Furthermore, at Uranus, grains
as close as $R_{syn}$ can escape, unlike at Jupiter. In
Fig.~\ref{fig:URA12}a, a spike at ($\Lpar = -0.03, r_L/R_p = 2$)
appears to be associated with the 2:1 inner Lorentz resonance that
approaches the Kepler limit at 2.04 $R_p$. Interestingly, this resonance
appears stronger for negative grains than for positive ones. At
Jupiter, two spikes in Fig.~\ref{fig:all}b distinguish the inner 2:1
vertical and radial resonances. Similarly, at Uranus, Fig.~\ref{fig:URA12}a hints
at an even stronger pairing of destabilizing 2:1 resonances, one curving downward
towards the planet as $|\Lpar|$ increases, and one arcing slightly upward.

\subsection{Neptune}
\begin{figure} [placement h]
\includegraphics [height = 2.1 in] {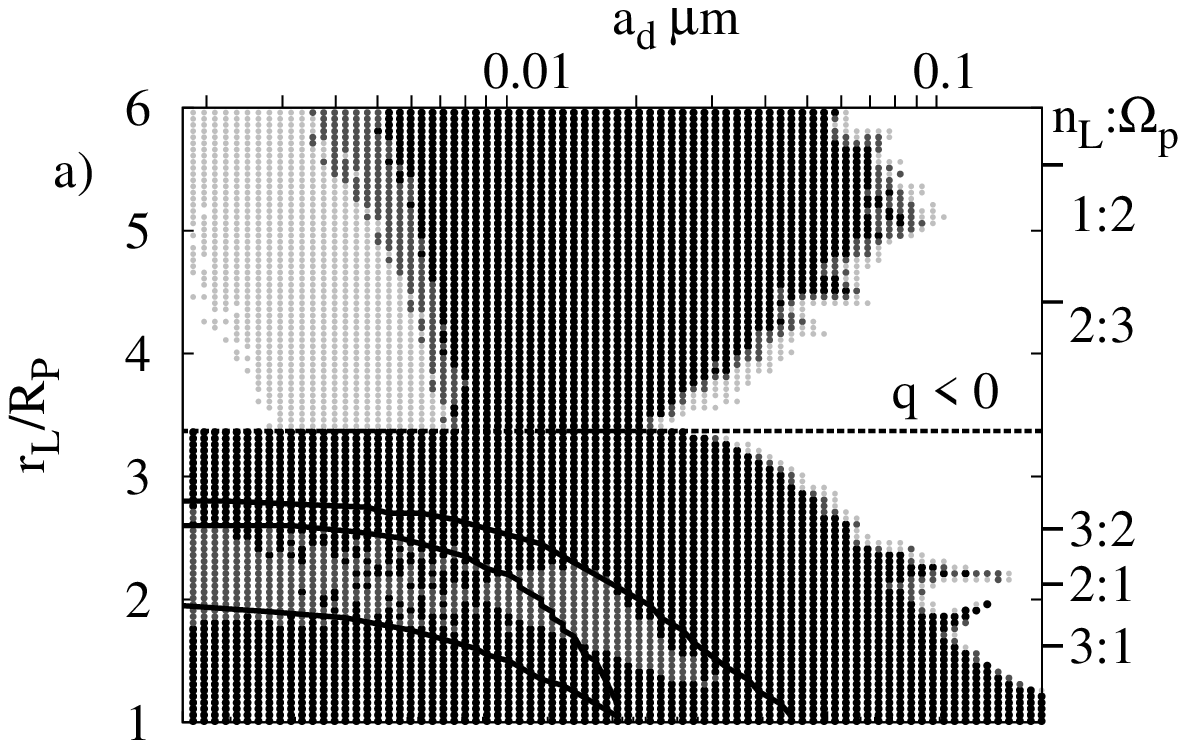}
\newline
\includegraphics [height = 2.1 in] {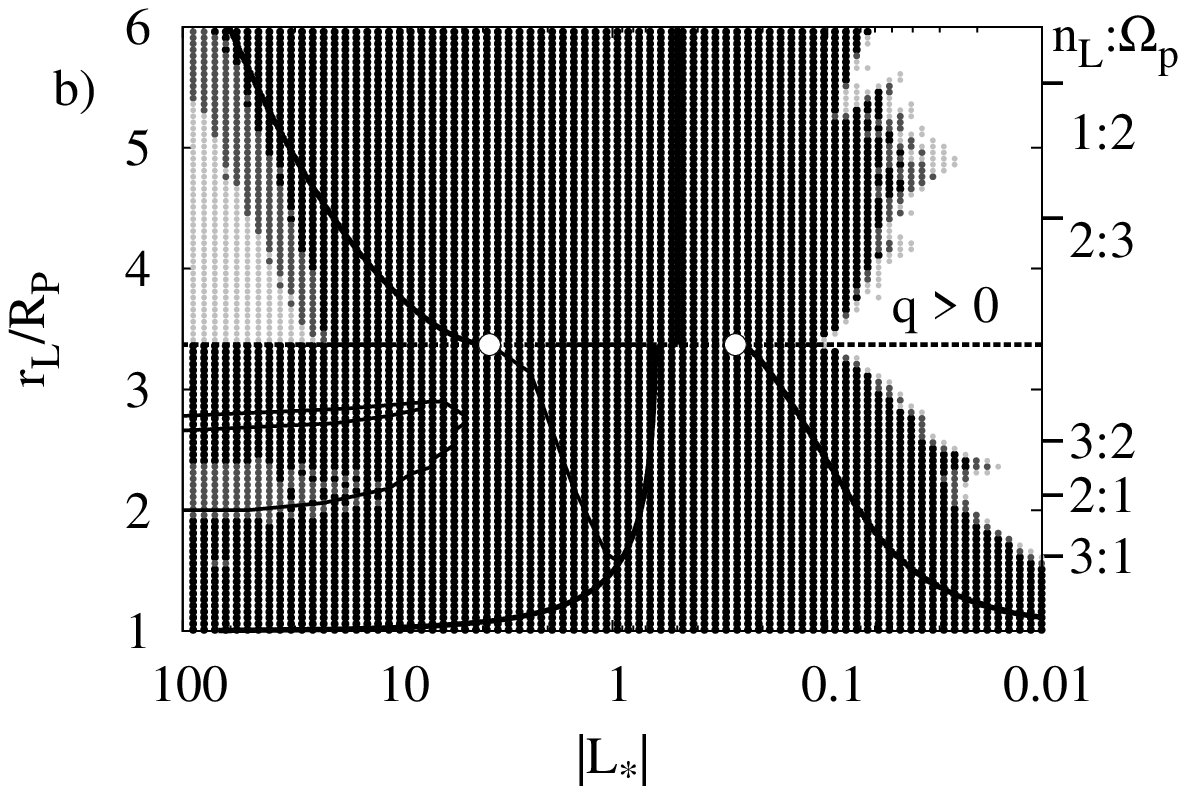}
\caption{Stability of Kepler-launched grains in Neptune's full
  magnetic field, for twelve equally-spaced azimuthal launch
  positions, and integrated over one year. The greyscale is as in
  Fig.~\ref{fig:URA12} with the darkest grey unstable for all launch
  longitudes and grains in the white areas surviving for all launch
  positions. As in previous figures, we superpose numerically-determined 
  stability results for an aligned dipole as solid curves, 
  and mark the local radial stability threshold at $R_{syn}$, where $\Lpar=2\pm\sqrt{3}$ with white circles.}
\label{fig:NEP12}
\end{figure}

We model Neptune's magnetic field configuration with data from
\citet{can91}. As for Uranus, above, the stability map includes the
effect of launch longitude on grain-orbit stability. And as with
Fig.~\ref{fig:URA12}, Fig.~\ref{fig:NEP12} indicates the number of 12
equally-spaced launch azimuths that survive a 1-year integration.

Figure~\ref{fig:NEP12}a maps the stability of negatively-charged dust
at Neptune, and includes a large region of escaping negative grains,
though this range of is slightly smaller at
Neptune than at Uranus (Fig.~\ref{fig:URA12}a). However, the escape
region for negative grains is still much more significant at Neptune than at Jupiter (see
Fig.~\ref{fig:all}b), and it too reaches $R_{syn}$. As at Uranus,
grain orbit stability on the Lorentz-dominated side of
Figs.~\ref{fig:NEP12}a and~\ref{fig:NEP12}b is strongly dependent on
the launch phase. Inside synchronous orbit, grains in much of the highly-charged Lorentz limit
are unstable, except for a small region around $r_{L}/R_p= 2.2$ for
both positive and negative grains, where stability varies significantly with
launch phase. This contrasts with Uranus, where all grains inside
$R_{syn}$ in the Lorentz regime were unstable. Again, however, the
instability at Neptune vastly exceeds that of Jupiter.

In the Kepler-dominated regime of Fig.~\ref{fig:NEP12}a, two spikes most likely associated with
the inner 2:1 Lorentz resonance feature prominently. Just as we saw at Uranus, in
Fig.~\ref{fig:NEP12}b, this inner 2:1 resonance appears to be weaker. For negative grains, the white stable zone to the right in Fig.~\ref{fig:NEP12}a reaches to higher
$\Lpar$ values (smaller grain radii) than we saw for Uranus (Fig.~\ref{fig:URA12}a). For
both planets, the dependence on azimuthal launch position for stability
is only important for $|\Lpar| >>1$. In the Kepler regime, grains move rapidly across magnetic field lines, and instabilities are effectively averaged over all launch phases. All evidence points to greater instability at Uranus than at Neptune. This is consistent with the Uranian dipole tilt of $59^{\circ}$ exceeding Neptune's $47^{\circ}$.
\section{Discussion}
In this paper, we have studied three main effects on charged particle motion: i) non-zero launch velocities from orbiting parent bodies ii) complex magnetic fields and iii) time-variable electric charges. We presented data from over 250,000 numerical integrations and compared our results to analytical theories, extending the important concept of Lorentz Resonances to arbitrary charge-to-mass ratios, and showed that effects of order $\Lpar^2$ can explain results that the linear theory of \citet{ham94} misses. 

Non-zero launch impulses (section 3.1), relative to the Kepler
flow have little effect on charged-grain dynamics and stability. Radial stability boundaries are only noticeably affected by an azimuthal kick. Vertical instability, by contrast, is affected by vertical impulses as might be expected, but also by azimuthal kicks which strengthen or weaken the magnetic mirror force. Finally, a radial kick barely affects dust grain motions at all.

When considering the stability of grains in the wide variety of planetary magnetic fields in the Solar System, we have shown in section 4 that dust grains with constant charges provide the maximum stability possible at each planet. The simplest magnetic field that we have considered, that of Saturn, is well-described as an untilted dipole, moderately offset to the north. This offset noticeably expands the vertical instability but has little discernable effect on radial motions. 
 
Jupiter's magnetic field is substantially more complex than Saturn's with a moderate tilt, a southward offset, and sizeable higher-order field coefficients. The tilted dipole strongly affects vertical stability boundaries, and the loss of axisymmetry powers Lorentz Resonances. These resonances act to destabilize dust particles, allowing even negative grains outside synchronous orbit to escape from Jupiter. Thus the Io plasma torus, in which grains are expected to have negative charges, is not an impermeable barrier to escape. The high-speed dust streams detected by Ulysses and Galileo near Jupiter likely originate in the Io torus. We extend Lorentz resonances from the Kepler regime by rewriting their frequencies in terms of the general radial, vertical, and azimuthal frequencies valid at all charge-to-mass ratios. This allows us to calculate the radial locations of Lorentz resonances as a function of $\Lpar$. We note strong correlations between zones of instability and the predicted locations of Lorentz resonances. Finally, our numerical simulations show that some resonances with strengths proportional to $\Lpar^2$ must be active. We show how to determine the frequencies and rough stengths of these high-order resonances.  

Our results for Jupiter are directly applicable to the Earth, which also has a magnetic field that is dominated by a moderately-tilted dipole. Two interesting differences, however, distinguish charged-particle motion at Earth from that at Jupiter. Firstly, due to its inverted magnetic dipole, the radial instability at Earth affects negative not positive charges. Similarly, positively-charged dust at Earth behaves as negatively-charged dust at Jupiter. Secondly, the Earth is very small compared to the size of its synchronous orbital distance, minimizing the tendency of the vertical instability to force grains to collide with Earth. Furthermore, due to the rapid decay with distance of higher-order magnetic field terms, the effect of Earth's full magnetic field on dust grain trajectories differs little from that of a simple tilted dipole.
 
Uranus and Neptune both have complex magnetic field configurations
which render aligned or even simple tilted dipolar models
insufficient. Both of these planets have substantial quadrupolar and
octupolar components, which act to destabilize both negative and positive grains
across the synchronous orbital distance, and over a far greater range of charge-to-mass ratios than at the other planets that we have studied. These distant planets highlight how increases in magnetic field complexity dramatically exacerbate dynamical instabilities. Future spacecraft missions will provide more detailed planetary magnetic field configurations than we have available today, but changes to the stability maps that we have provided here for constant-charge dust grains are likely to be modest.

Relaxing the assumption of constant charge-to-mass ratios also leads to a substantial increase in the range of dust-grain sizes that are destabilized. Results are highly model-dependent, and for simplicity we adopted a sparse plasma with constant spatial density and photoelectric charging. The time-variable charging currents on a dust grain due to passage through the planetary shadow significantly expand the range of grain sizes that are globally
unstable, particularly inside synchronous orbit. Larger dust grains
respond rapidly to changes in the charging environment and hence
stability is determined by comparing the destabilizing timescale of
variable Lorentz forces with the orbital precession time (\citealt{hb91,hk08}). For our nominal model, this increases the threshold radius for destabilized grains by more than an order of magnitude. For smaller grains, by contrast, charging is slow with the result that different stability curves apply at different times, expanding the zone of instability by an order of magnitude over that expected for a constant charge. The sparse plasma that we have adopted here is appropriate for the dusty main ring and gossamer rings at Jupiter. We find that the removal of dust at Jupiter is dominated by the basic dipolar radial instability for positive grains, substantially extended to both larger and smaller particles by the effects of variable charging. We leave a full study of the dependence of these effects on plasma parameters for a future study. 
\section{Acknowledgements}
We thank Leslie Schaffer and a second anonymous referee for helpful reviews. This work was supported by the NASA Outer Planets research program.

\end{document}